\documentclass[11pt]{article}
\usepackage[a4paper,margin=2.5cm]{geometry}
\usepackage[T1]{fontenc}
\usepackage[utf8]{inputenc}
\usepackage{lmodern}
\usepackage{amsmath,amssymb}
\usepackage{graphicx}
\usepackage{xcolor}
\usepackage{siunitx}
\usepackage[superscript,biblabel]{cite}
\usepackage{caption}
\usepackage{booktabs}
\usepackage{algpseudocode}
\PassOptionsToPackage{obeyspaces}{url}
\usepackage{xurl}
\usepackage[hidelinks]{hyperref}
\usepackage{setspace}
\usepackage{flafter}
\title{\bfseries Inverse design of large-scale freeform meta-optics by breaking the memory wall of full-wave simulation}
\author{Hyoseok Park and Yeonsang Park\thanks{Corresponding author: yeonsang.park@cnu.ac.kr}\\[2pt]
\small Department of Physics, Chungnam National University, Daejeon, Republic of Korea}
\date{}

\newcommand{\micron}{\ensuremath{\mu\mathrm{m}}}
\newcommand{\suppnote}[1]{Supplementary Note~\ref{si:#1}}
\newcommand{\suppfig}[1]{Supplementary Fig.~\ref{fig:#1}}
\newcommand{\supptab}[1]{Supplementary Table~\ref{tab:#1}}
\begin{document}
\maketitle
\begin{abstract}
Gradient-based design of large optical surfaces hits a memory wall: a stored-history time-domain adjoint needs $24N_{t}$ bytes per cell for $N_{t}$ steps, and every field update is bound by memory bandwidth. Here we present a validated framework that removes this wall and makes globally optimized freeform meta-optics practical. Its finite-difference time-domain adjoint reconstructs interior fields backwards from two recorded planes, so volume memory no longer grows with $N_{t}$, and fused kernels raise memory-bandwidth use on an H200 GPU from 6.4\% to a remarkable 56\% of peak. Overlapping tiles, coupled by angular-spectrum propagation and distributed over GPUs, keep tile memory independent of the aperture. Within hours on one GPU node, we optimize, against one global objective, every density variable of single-layer metasurfaces 100 to \SI{200}{\micro\metre} across, and design a nine-wavelength lens, a color hologram and a polarization-switched hologram. Confirmed by an independent solver within 1.6\%, they substantially outperform meta-atom designs, reaching 1.64, 1.65 and 1.24 times their objectives.
\end{abstract}

Adjoint methods have made gradient-based design a working tool across computational science, including photonic device design\cite{cpc_jensen2011,cpc_molesky2018}, because one forward and one reverse solve give the derivative with respect to every design variable. In the time domain, the reverse solve consumes the forward fields in reverse order, so memory, not arithmetic, decides how large a design can be. Optical metasurfaces push this limit to its extreme: a device a fifth of a millimeter across already holds more than $10^{10}$ grid cells.

\begin{figure}[p]
\centering
\includegraphics[width=\linewidth]{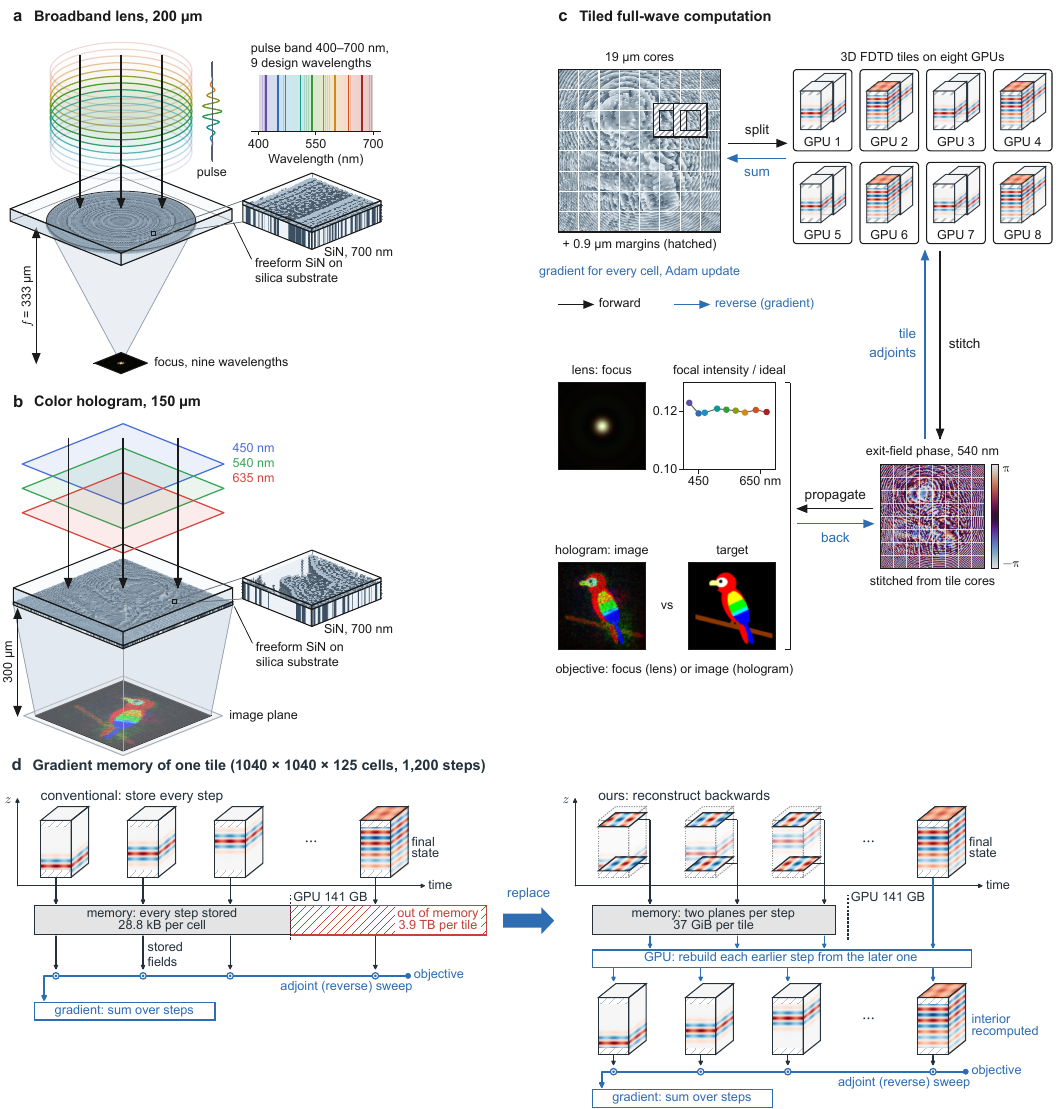}
\caption{\textbf{Overlapping three-dimensional FDTD tiles compute the freeform layer, transposes of every stage return its gradient, and reverse reconstruction fits each tile gradient on one GPU.} (a) A broadband pulse from the silica substrate passes the freeform silicon nitride layer (height exaggerated) of the \SI{200}{\micro\metre} lens, which focuses nine wavelengths from 420 to \SI{670}{\nano\metre} at $f=\SI{333}{\micro\metre}$. (b) The \SI{150}{\micro\metre} color hologram forms its image \SI{300}{\micro\metre} away at 450, 540 and \SI{635}{\nano\metre}. The insets in (a) and (b) show \SI{4}{\micro\metre} of each layer on its \SI{10}{\nano\metre} grid. (c) The layer is cut into windows with \SI{19}{\micro\metre} cores and \SI{0.9}{\micro\metre} margins (hatched, drawn enlarged), solved as tiles on eight GPUs, and the tile cores form the exit field, shown for the hologram as its phase at \SI{540}{\nano\metre}, which is propagated to the focus or the image and scored by the objective $J$. The reverse chain (blue) applies the transposes in the opposite order and returns the gradient of $J$ for every cell of the layer. (d) Gradient memory of one production tile ($1040\times1040\times125$ cells, 1,200 steps). The adjoint sweep sums the products of the adjoint and forward fields over all steps. A stored-history adjoint (left) keeps the forward field of every step and needs 3.9~TB, beyond the 141~GB of one H200, whereas reverse reconstruction (right) records only the two planes that bound the absorbers at every step and rebuilds the interior backwards in time, which needs 37~GiB.}
\label{fig:framework}
\end{figure}

Metasurfaces shape the phase, amplitude and polarization of light with one patterned layer of subwavelength features\cite{ncs_yu2011,ncs_yu2014,ncs_arbabi2015,ncs_khorasaninejad2016,ncs_kuznetsov2024}. Most large metasurfaces are designed by the meta-atom method: a library of periodic unit cells is computed once, a target phase profile is sampled on a lattice, and each site receives the library element whose response is closest to the target\cite{ncs_chen2020review,ncs_park2024}. The method rests on the locally periodic approximation, in which each element scatters as if all its neighbors were identical copies of itself\cite{cpc_pestourie2018,ncs_lpa2019}. The approximation holds where the target phase varies slowly across several lattice periods and each element must meet one condition. It weakens where the phase changes by a large fraction of $2\pi$ between neighbors, where one element must set independent phases at several wavelengths or polarizations, and where the target lacks a smooth envelope. Adjoint inverse design removes these assumptions by optimizing the material distribution against the optical function itself\cite{cpc_jensen2011,cpc_molesky2018,ncs_christiansen2021,ncs_li2022review}, but the cost of full-wave solutions has confined it to small apertures or to reduced design spaces\cite{ncs_roberts2023,ncs_kang2024}.

The obstacle is the size of the full-wave problem. A single-layer metasurface \SI{200}{\micro\metre} across, meshed at \SI{20}{\nano\metre} over a vertical extent of \SI{2.5}{\micro\metre}, contains $1.25\times10^{10}$ Yee cells. Their six single-precision field components alone occupy 279~GiB, about twice the 141~GB of an H200 GPU, and a time-domain adjoint needs this state at every step of its reverse sweep. Simulating the path to a focal plane \SI{333}{\micro\metre} away would enlarge the domain to about $1.7\times10^{12}$ cells and lengthen the run from the 1,200 steps of a near-field calculation to about $3.2\times10^{4}$ steps. At a fixed numerical aperture (NA) the cell updates of such a direct calculation grow with the fourth power of the aperture diameter $D$, whereas a near-field calculation followed by angular-spectrum propagation grows with $D^{2}$. Speed is limited by memory as well: each time step reads and writes every field component but does little arithmetic on it, so the step time is set by memory bandwidth. Capacity and bandwidth form one memory wall\cite{ncs_gholami2024}, and it confronts every explicit time-domain adjoint.

Earlier large-area methods reduce this cost by restricting the model or the design. Locally periodic topology optimization evaluates each surface cell with periodic boundaries\cite{ncs_lpa2019}, and overlapping-domain optimization replaces them with padded local solutions in two dimensions\cite{cpc_lin2019,ncs_zhelyeznyakov2023}. Three-dimensional adjoints have optimized post widths\cite{ncs_mansouree2021}, the positions and radii of scatterers\cite{ncs_skarda2022} and millimeter-scale lenses built from repeated axisymmetric zones\cite{ncs_zda2025}; such zoned designs repeat each radial zone around the axis, whereas here every density variable of the aperture is optimized jointly against one global objective. Augmented partial factorization avoids volumetric solves for many inputs\cite{ncs_apf2022,ncs_apfinverse2024}, and distributed CPU and GPU time-domain solvers have simulated and designed large structures\cite{ncs_hammond2022,ncs_hughes2021,cpc_mahlau2024,cpc_mahlau2026joss}. Each approach exchanges design freedom, electromagnetic coupling or dimensionality for cost (\suppfig{landscape}a). We target a density that varies freely over the whole aperture of a three-dimensional device, coupled by full-wave fields without local periodicity.

Here we remove the memory barrier of three-dimensional full-wave adjoint optimization for such designs. The reverse sweep reconstructs the interior fields backwards from the final state and records only the two planes that bound the absorbers\cite{ncs_anderson2012,ncs_yang2016checkpoint,ncs_lu2021patent,ncs_tang2023timereversal,cpc_mahlau2024,cpc_mahlau2026joss}, and fused kernels keep a reverse step at 5.9~ms, 1.4 times the 4.2~ms of a forward step. We apply this adjoint to apertures that hold $10^{10}$ Yee cells. For a production tile of $1040\times1040\times125$ cells (a \SI{20.8}{\micro\metre} window on a \SI{20}{\nano\metre} mesh) and 1,200 steps, the gradient memory falls dramatically, from 3.9~TB to 37~GiB on one GPU, for three or nine recorded wavelengths alike. We implement this adjoint in TorchFDTD\cite{torchfdtdsoftware}, our GPU finite-difference time-domain (FDTD) solver, in which each update and its transpose run as one fused kernel that reads each field once per step. Around this adjoint, we assemble a full aperture from overlapping tiles whose window gradients are summed, couple the tiles to the observation plane by a differentiable vector angular spectrum, and distribute them over GPUs (Fig.~\ref{fig:framework}). The tiles exchange no data during a solve, so the tile memory per GPU stays remarkably flat: in an aperture benchmark the tile solves peak at 36.7 to 38.2~GiB per GPU at every aperture.

We design three devices 100 to \SI{200}{\micro\metre} across, each a planar density extruded through a \SI{700}{\nano\metre} silicon nitride layer: a lens that focuses nine visible wavelengths, a hologram that projects a full-color image with one wavelength per color channel, and a hologram that projects one of two words depending on the input polarization. Remarkably, each design takes only 4.7 to 10.6~h of optimization on one GPU node, or 37.7 to 72.8 GPU hours. A direct time-domain calculation of the lens would need about $1.6\times10^{5}$ GPU hours, more than three orders of magnitude more. The three designs outperform meta-atom designs, reaching 1.64, 1.65 and 1.24 times their objectives. All three are scored with one full-wave evaluator, and single-domain solutions and Tidy3D confirm the scores over the whole aperture. The advance is computational: an exact transpose of every stage, from the tiles through angular-spectrum propagation, carries a three-dimensional time-domain adjoint to $10^{10}$ cells. To our knowledge, these are the first three-dimensional full-wave inverse designs of freeform metasurfaces 100 to \SI{200}{\micro\metre} across in which the whole aperture is optimized as one layer, without meta-atom-scale locally periodic unit cells or periodically repeated zones, the lens with D4 symmetry and the holograms without symmetry.

\section*{Results}

\subsection*{Full-aperture gradients from independent tiles}
Every stage between the latent density and the objective has a transpose, so we assemble the gradient of a whole aperture from independent three-dimensional FDTD tiles (Fig.~\ref{fig:framework}a--c). In the forward pass, the design is a density on a \SI{10}{\nano\metre} grid. A filter smooths it and a projection pushes it toward 0 or 1, which gives the pattern of the \SI{700}{\nano\metre} silicon nitride layer. We then cut the aperture into windows, each a square core plus a margin on every side, and simulate each window as a three-dimensional FDTD tile illuminated from the substrate by one broadband pulse (\suppfig{tilegeom}). We call the margin the overlap, because it reaches into the neighboring cores. The tile cores form one global exit plane $u=\sum_jR_jC_ju_j$, which the angular spectrum $P_z$ carries to the observation planes. There the objective $J$ measures focusing at nine wavelengths for the lens (E1), a color image at 450, 540 and \SI{635}{\nano\metre} for the color hologram (E2) and two images selected by the input polarization for the polarization-switched hologram (E3). The reverse sweep applies $P_z^{\mathsf{T}}$, the restriction $C_j^{\mathsf{T}}R_j^{\mathsf{T}}$ of the cotangent to the tile cores, the tile adjoints and the sum $\sum_jS_j^{\mathsf{T}}$ of whole-window gradients, margins included. One pass returns the derivative with respect to all $7.85\times10^{7}$ latent variables of the lens, and up to $2.25\times10^{8}$ for the holograms (Methods).

The permittivity gradient sums the product of the adjoint and forward fields over all steps, so the reverse sweep needs the forward fields in reverse order, and storing all $N_{t}$ of them exhausts the memory of a conventional time-domain adjoint. We regenerate them instead (Fig.~\ref{fig:framework}d). The interior of a tile is lossless, so the Yee update runs backwards there, recovering each earlier field from the later one while the material cotangent accumulates. The absorbers are lossy, so the fields on the two planes that bound them are recorded at every step, and the tile memory is set by its volume and surface, not by the number of steps. The reverse sweep is the exact transpose of the discrete update, so it returns the derivative of the discretized model to round-off.

\begin{figure}[tbp]
\centering
\includegraphics[width=1.0\linewidth]{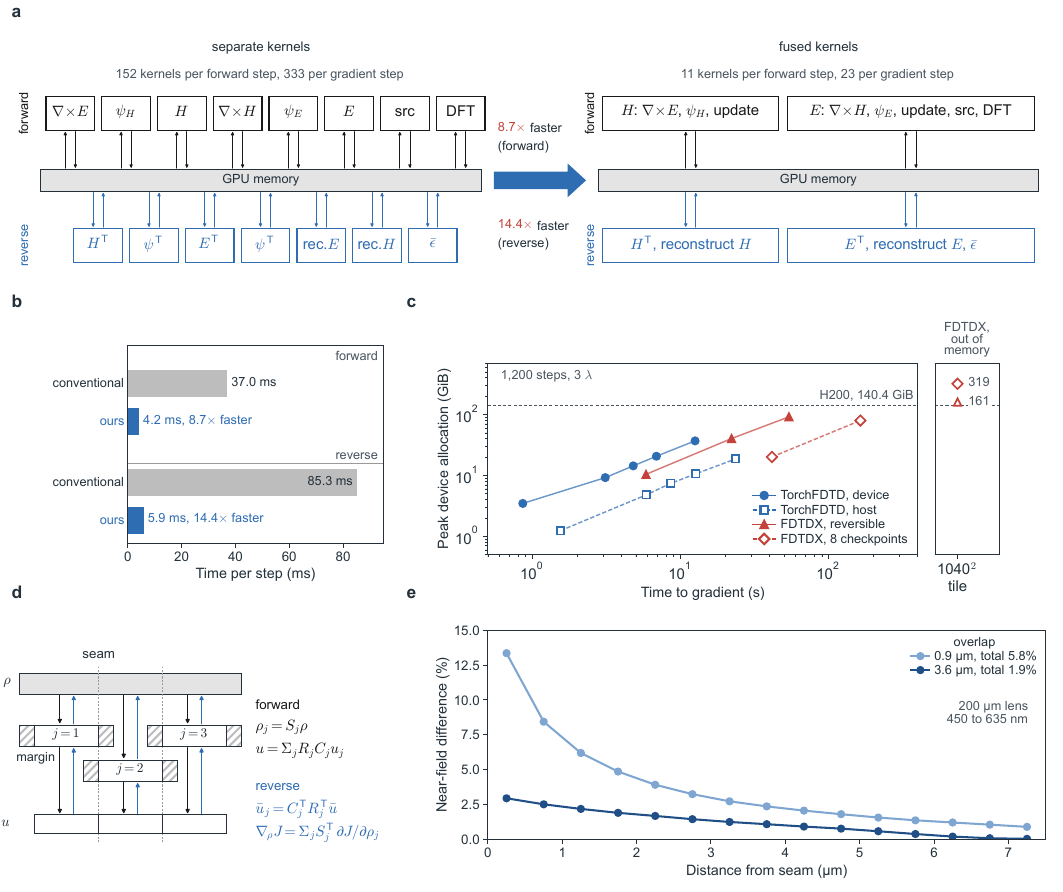}
\caption{\textbf{Fused kernels and overlapping tiles make each full-aperture gradient fast, small and accurate.} (a) Separate kernels read and write the fields in GPU memory for every operation, whereas a fused kernel reads each field once per update. (b) Time per forward and reverse step of a production tile on one H200, with separate kernels (conventional) and with fused kernels that reconstruct only the design layers (ours). (c) Peak device memory against time to gradient for one tile, TorchFDTD and FDTDX. (d) The forward assembly keeps the tile cores, and the reverse sweep sums the gradients of whole windows, margins included. (e) Near-field difference of the production tiles of the \SI{200}{\micro\metre} meta-atom lens against one TorchFDTD domain, against the distance from the nearest seam.}
\label{fig:methods}
\end{figure}

Reconstruction adds work to every reverse step, so its speed matters. Fused kernels, with the reconstruction restricted to the design layers, cut the cost of a tile step by a factor of 9 to 14. We write each update and its transpose as fused CUDA kernels that read each field once per step (Fig.~\ref{fig:methods}a), and we restrict the reconstruction to the layers that carry the design. For the production tile of $1040\times1040\times125=1.35\times10^{8}$ cells, the forward and reverse steps take 4.2 and 5.9~ms, 8.7 and a full 14.4 times faster than the 37.0 and 85.3~ms of the same update executed as separate PyTorch tensor operations, with bitwise identical spectra (Fig.~\ref{fig:methods}b). They reach 56\% of the 4.8~TB/s peak memory bandwidth of an H200 (about 2.7~TB/s), against 6.4\% (0.31~TB/s) for the separate operations (\supptab{fused} and \suppnote{kernels}). Together with the reverse reconstruction, the gradient of a production tile of 1,200 steps takes 12.5~s and peaks at 37.0~GiB on one H200. Neither gradient mode of FDTDX, an independent GPU FDTD solver with automatic differentiation\cite{cpc_mahlau2024,cpc_mahlau2026joss}, could hold this tile on the 141~GB device; they would need 160.5~GiB in the reversible mode and 319~GiB with eight checkpoints (Fig.~\ref{fig:methods}c and \suppnote{xsolver}).

The solutions of the overlapping tiles are stitched together, core by core, into one exit field of the whole aperture, and the gradient is the exact derivative of this stitched field. A density cell in an overlap lies in the windows of up to four tiles, so its gradient sums the cotangents from every window that contains it (Fig.~\ref{fig:methods}d). The exterior is homogeneous, so a vector angular spectrum computed by fast Fourier transforms carries the exit plane to the observation planes, and its transpose returns the exit-plane cotangent. Within one pass the tiles exchange nothing, so we distribute the 32 to 64 tile solutions of one gradient over the GPUs of a node (Methods). Adding GPUs shortens a step without changing its result.

We test the whole chain, from latent density to objective, by directional central differences on a \SI{12}{\micro\metre} aperture with the production mesh and pupil sampling. In eight cases with three test objectives and gradient-aligned or random directions, the central differences match the adjoint derivative to $1.5\times10^{-4}$ or better (\supptab{gradverify}). The gradient of one tile also agrees with that of FDTDX, with a cosine similarity above 0.999 (\suppfig{fdtdxadj}). On a \SI{30}{\micro\metre} test with a coarser \SI{50}{\nano\metre} mesh, the tiled gradient converges to the full-domain gradient as the overlap grows (\suppnote{overlap}), and on the \SI{200}{\micro\metre} meta-atom lens the near-field difference from one domain is concentrated at the seams (Fig.~\ref{fig:methods}e). We run the optimizations with \SI{0.9}{\micro\metre} overlaps and evaluate every objective reported for the three designs at full wave over the whole aperture, with tiles of \SI{15}{\micro\metre} cores and \SI{3.6}{\micro\metre} overlaps.

\subsection*{One freeform layer focuses nine wavelengths}
We design a lens \SI{200}{\micro\metre} in diameter with a focal length of \SI{333.3}{\micro\metre}, so $R/f=0.30$, to focus nine wavelengths from 420 to \SI{670}{\nano\metre} in both polarizations onto one plane. Its figure of merit at each wavelength, averaged over the two polarizations, is the on-axis intensity at the common focus relative to that of an ideal lossless lens of the same aperture, which equals the Strehl ratio times the transmission. We start from a gray density, share the variables among the symmetric octants of the aperture and maximize a soft minimum of this quantity over wavelength, as in the maximin formulations of earlier broadband lens designs\cite{ncs_chung2020,ncs_li2022natcommun}.

\begin{figure}[tbp]
\centering
\includegraphics[width=0.93\linewidth]{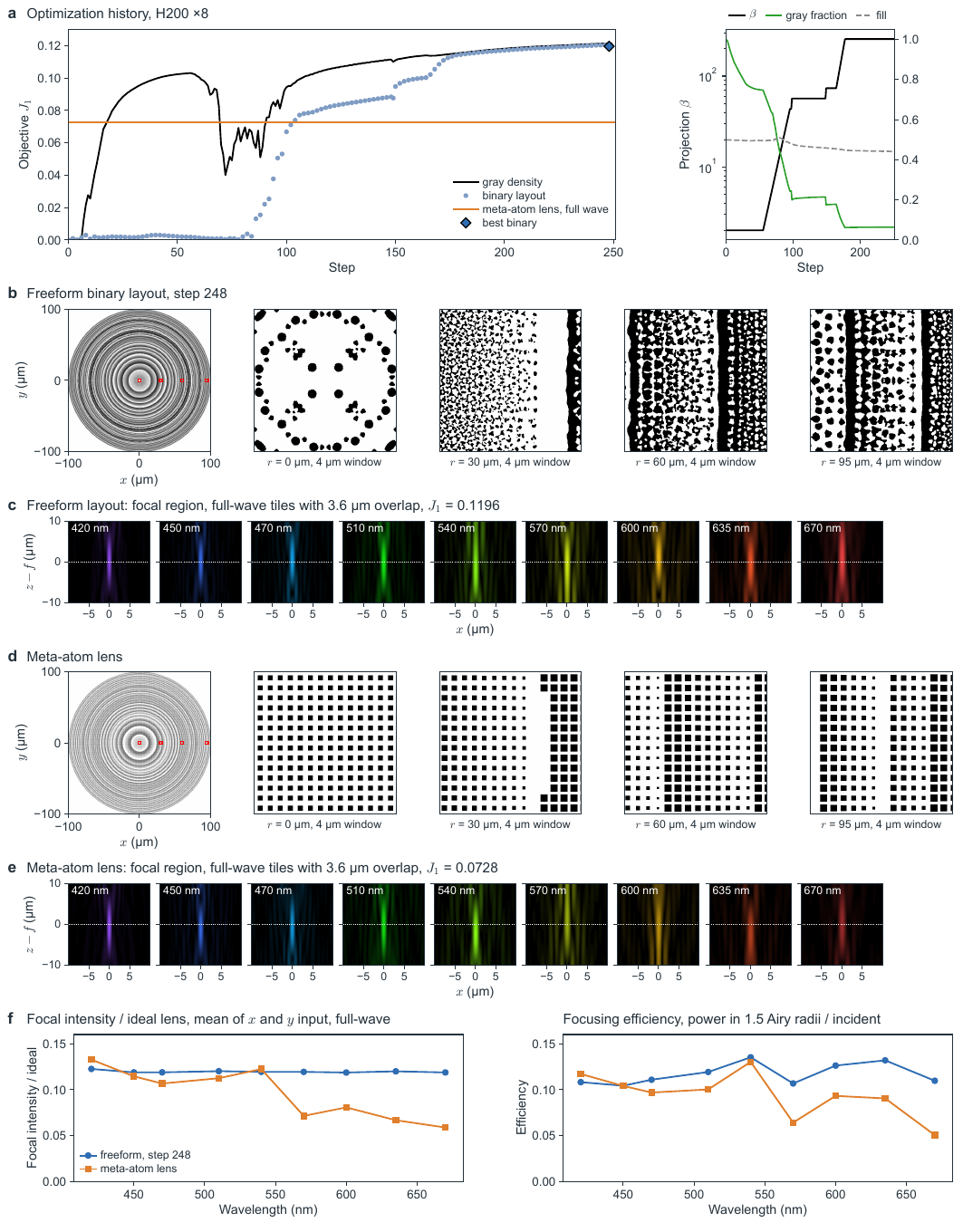}
\caption{\textbf{One freeform layer focuses nine wavelengths with an evenly allocated chromatic error.} \SI{200}{\micro\metre} broadband lens. (a) Objective $J_{1}$ of the gray density (black line) and of its thresholded binary layout (points) against step, recorded with \SI{0.9}{\micro\metre} tile overlaps, with the best binary layout (diamond) and the meta-atom lens (orange line) scored with \SI{3.6}{\micro\metre} overlaps; right, projection sharpness $\beta$, gray fraction and fill of the density. (b) Best binary layout, step 248, over the aperture and in four \SI{4}{\micro\metre} windows at radii of 0, 30, 60 and \SI{95}{\micro\metre} (red squares). (c) Focal region $S_{z}(x,y=0,z)$ of that layout for $x$ input at the nine wavelengths, divided by the focal $S_{z}$ of the ideal lens at the same wavelength. (d) Meta-atom lens from the post library on a \SI{280}{\nano\metre} lattice, with the same windows. (e) Focal region of the meta-atom lens as in (c). Each wavelength is shown on a scale from black to its spectral color, and all panels of (c) and (e) share one intensity scale from 0 to 0.138. (f) Focal intensity $S$ relative to the ideal lens, averaged over $x$ and $y$ input, and focusing efficiency, the power within 1.5 Airy radii over the incident power, at each wavelength for the freeform (blue) and meta-atom (orange) lenses.}
\label{fig:achromat}
\end{figure}

The meta-atom lens follows dispersion engineering\cite{ncs_chen2018,ncs_wang2018,ncs_shrestha2018,ncs_chen2023disp}: from a library of square silicon nitride posts on a \SI{280}{\nano\metre} lattice, each site receives the post whose phase spectrum best matches the ideal lens phase plus a reference phase per wavelength shared by the whole aperture (\suppnote{baselines}). Scored by full-wave tiles with \SI{3.6}{\micro\metre} overlaps, the best binary freeform layout reaches $J_{1}=0.1196$, 1.64 times the 0.0728 of the meta-atom lens (Fig.~\ref{fig:achromat}a, Extended Data Table~\ref{tab:gains} and Methods). As the projection sharpness $\beta$ rises, the gray density is driven toward 0 and 1, and the score of the binary layout climbs from near zero to about 0.7\% below the gray objective (Fig.~\ref{fig:achromat}a). The freeform layout forms curved, aperiodic features that change with radius, whereas the meta-atom lens is a lattice of square posts (Fig.~\ref{fig:achromat}b,d).

Crucially, the freeform lens spreads its chromatic error evenly across the band, whereas the meta-atom lens concentrates it at the red end, where its focus fades (Fig.~\ref{fig:achromat}c,e). Averaged over $x$ and $y$ input, the focal intensity of the freeform layout lies between 0.119 and 0.123 of the ideal lens at all nine wavelengths, whereas that of the meta-atom lens falls from 0.133 at \SI{420}{\nano\metre} to 0.059 at \SI{670}{\nano\metre} (Fig.~\ref{fig:achromat}f). The weakest wavelength of the freeform lens still has twice the focal intensity of the weakest wavelength of the meta-atom lens, 0.119 against 0.059. The freeform focus keeps a full width at half maximum of 1.01 to 1.10 times $\lambda/2\mathrm{NA}$, and its peak lies within 0.31 depths of focus of $f$ at every design wavelength (\suppnote{objectives}). Its focusing efficiency is 0.104 to 0.135 within 1.5 Airy radii and 0.099 to 0.130 within the disk of three focal full widths used for earlier lenses\cite{ncs_chen2018,ncs_zda2025} (\suppnote{objectives}). The flat response reflects how the layer handles group delay: the lens asks for \SI{49}{\femto\second} of delay between axis and rim, and the freeform layer shares the delay it imposes evenly among the nine wavelengths.

\subsection*{Freeform holograms separate colors and polarizations}
\begin{figure}[tbp]
\centering
\includegraphics[width=\linewidth]{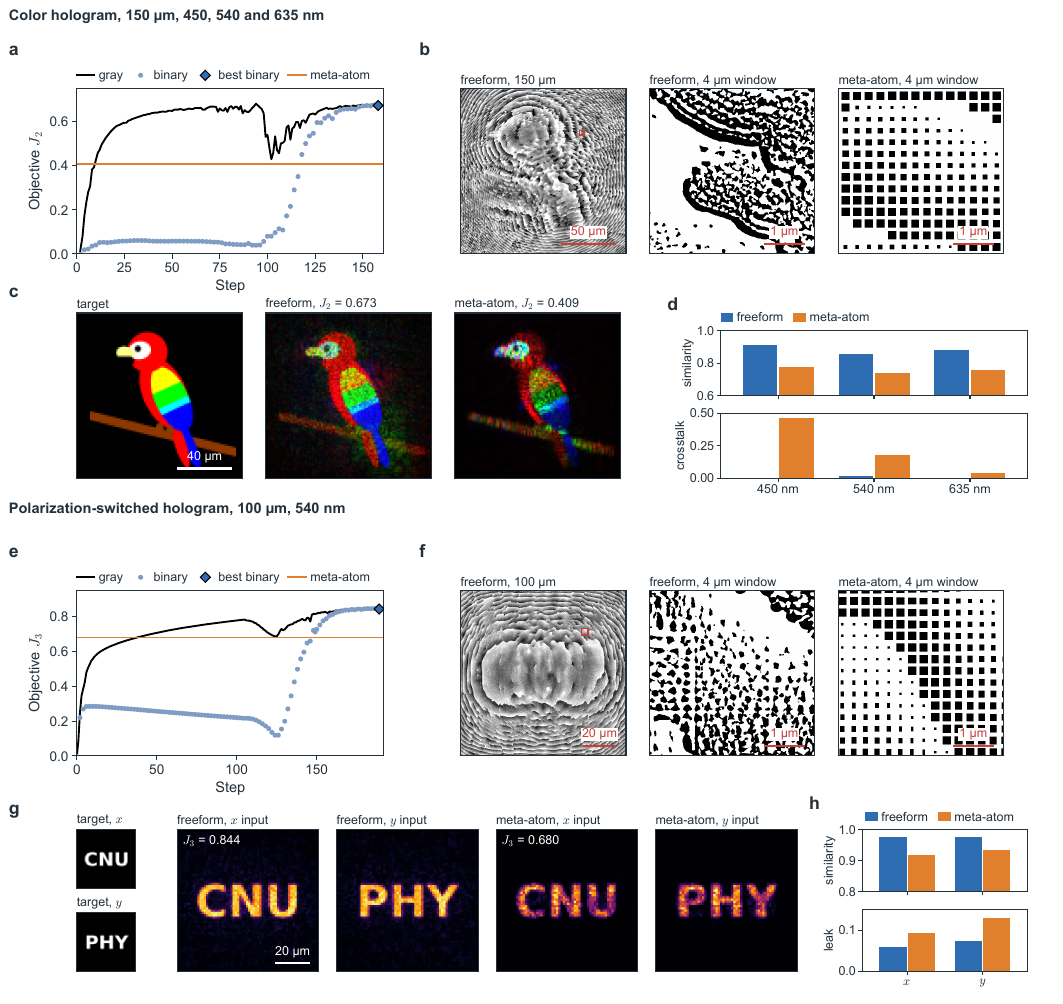}
\caption{\textbf{Freeform layers separate the color and polarization channels that meta-atom holograms mix.}
(a) Objective $J_{2}$ of the color hologram during optimization, for the gray density and the binary layouts, with the best binary layout and the meta-atom hologram at full wave.
(b) Best binary freeform layout over the \SI{150}{\micro\metre} aperture, one \SI{4}{\micro\metre} window on its \SI{10}{\nano\metre} grid and the same window of the meta-atom layout.
(c) Target and images of the freeform ($J_{2}=0.673$) and meta-atom ($J_{2}=0.409$) holograms \SI{300}{\micro\metre} behind the aperture, with red, green and blue at 635, 540 and \SI{450}{\nano\metre}, each image normalized to its own brightness.
(d) Similarity of each channel to its own target and crosstalk excess, summed over the other channels (Methods).
(e) Objective $J_{3}$ of the polarization-switched hologram during optimization.
(f) Freeform layout over the \SI{100}{\micro\metre} aperture, one \SI{4}{\micro\metre} window and the same window of the meta-atom layout.
(g) Target words and images at \SI{540}{\nano\metre}, \SI{200}{\micro\metre} behind the aperture, for $x$ and $y$ input, of the freeform ($J_{3}=0.844$) and meta-atom ($J_{3}=0.680$) holograms, each image normalized to its own maximum.
(h) Similarity of each image to its own word and leak into the letters of the other word.
All scores are full-wave, from tiles with \SI{3.6}{\micro\metre} overlaps; the histories in (a) and (e) were recorded with the \SI{0.9}{\micro\metre} overlaps of the optimization.}
\label{fig:holograms}
\end{figure}

A freeform layer separates color channels that a meta-atom library mixes. We next design a color hologram, a \SI{150}{\micro\metre} square aperture that should project, under $x$-polarized illumination, a procedurally generated full-color image of a parrot into a plane \SI{300}{\micro\metre} behind it. Its red, green and blue channels, at 635, 540 and \SI{450}{\nano\metre}, are the linear intensities of a $96\times96$-pixel image \SI{120}{\micro\metre} wide, and each channel should appear only at its own wavelength. The objective rewards, at each wavelength, the efficiency into the image window times the similarity of the image to its own channel, penalizes similarity to the other channels and takes a soft minimum over the three wavelengths (Methods). The meta-atom counterpart starts from a Gerchberg--Saxton phase for each channel\cite{ncs_gerchberg1972}, selects at each site the post whose three phases best match the three targets together\cite{ncs_shi2018}, and iterates the loop with the library transmissions (\suppnote{baselines}). The binary freeform layout reaches $J_{2}=0.673$, 1.65 times the 0.409 of the meta-atom hologram (Fig.~\ref{fig:holograms}a). Its pattern of curved bands and irregular islands has no lattice, whereas the meta-atom hologram places square posts of varying width on the \SI{280}{\nano\metre} lattice (Fig.~\ref{fig:holograms}b). The similarity of each channel to its own target rises from 0.739--0.776 to 0.856--0.911 (Fig.~\ref{fig:holograms}c,d and \suppnote{objectives}). The gain comes mainly from color separation: across the three channels, the largest crosstalk excess falls by a striking factor of 24, from 0.464 for the meta-atom hologram to 0.019 or below. Each wavelength now forms only its own image.

One freeform layer also switches between two images with the input polarization. A \SI{100}{\micro\metre} square aperture should project the word CNU under $x$-polarized and the word PHY under $y$-polarized illumination at \SI{540}{\nano\metre}. The images are the total $S_{z}$, without an analyzer, on $64\times64$ pixels of \SI{1.25}{\micro\metre} in a plane \SI{200}{\micro\metre} behind the aperture. The objective takes the form of that of the color hologram, with the two inputs in place of the wavelengths. The meta-atom counterpart selects at each site the rectangular post whose $x$ and $y$ transmissions best match two Gerchberg--Saxton phases and iterates the loop with the library (\suppnote{baselines}). The binary freeform layout reaches $J_{3}=0.844$, 1.24 times the 0.680 of the meta-atom hologram (Fig.~\ref{fig:holograms}e), again with an aperiodic pattern in place of a lattice of rectangular posts (Fig.~\ref{fig:holograms}f). Its similarity to its own word is 0.975 for both inputs, against 0.920 and 0.936. Less light leaks into the wrong word: the power in the letter pixels of the other word falls from 0.093 and 0.130 to 0.059 and 0.073 of the power in the letters of the intended word (Fig.~\ref{fig:holograms}g,h and \suppnote{objectives}).

\subsection*{Where meta-atom design falls short}
The meta-atom picture fails first where neighboring elements differ strongly. The phase gradient of the lens grows with radius, and at the rim the phase advances by $2\pi$ over 5.2 lattice periods at \SI{420}{\nano\metre} and 8.3 periods at \SI{670}{\nano\metre}. A hologram phase has no smooth envelope: iterative phases of an image are speckled and contain vortices, around which the phase winds by $2\pi$ within a few lattice periods anywhere in the aperture. Over the lattice sites of the meta-atom layouts, the 95th percentile of the phase step between neighbors is 3.05, 2.79 and 2.08~rad per period for E1, E2 and E3 (Extended Data Table~\ref{tab:gains}). In these three examples, the freeform gain tracks these steps: the two layouts with steps near 3~rad gain 1.64 and 1.65, and the layout with 2.08~rad gains 1.24. The locally periodic model tracks the same order, over-predicting the full-wave objective of the meta-atom lens and color hologram by 7\% and 11\% while matching that of the meta-atom polarization-switched hologram to 0.1\%. Where phases step by large amounts, the design model itself therefore needs to be full-wave.

The picture also fails where one element must serve several functions. A square post has one free parameter, its width, whereas the color hologram asks each site for three independent phases at 450, 540 and \SI{635}{\nano\metre}. As the width changes, the three phases move together along one path, so most combinations of three phases are out of reach. Dispersion engineering reduces the nine phases of the lens at each site to a phase and a group delay\cite{ncs_shrestha2018,ncs_chen2020review}. Light from the rim of the lens travels \SI{14.7}{\micro\metre} farther to the focus than light on the axis, a group delay of \SI{49}{\femto\second}. Across a library of posts, holes and crosses in the same layer, the group delay spreads by a median of 2.45~fs, 5\% of what the lens needs (\suppnote{baselines}). The rectangular posts of the polarization-switched hologram reach a retardance of 1.08~rad, a third of the $\pi$ of a half-wave plate. In each case one element must meet several conditions with one or two parameters; each \SI{280}{\nano\metre} cell of the freeform layer instead contains 784 latent pixels, and because neighboring cells couple, the layer can spread a wavelength-dependent phase over several cells. Coupling becomes a resource.

\subsection*{Single domains and an independent solver confirm the designs}
\begin{figure}[tbp]
\centering
\includegraphics[width=\linewidth]{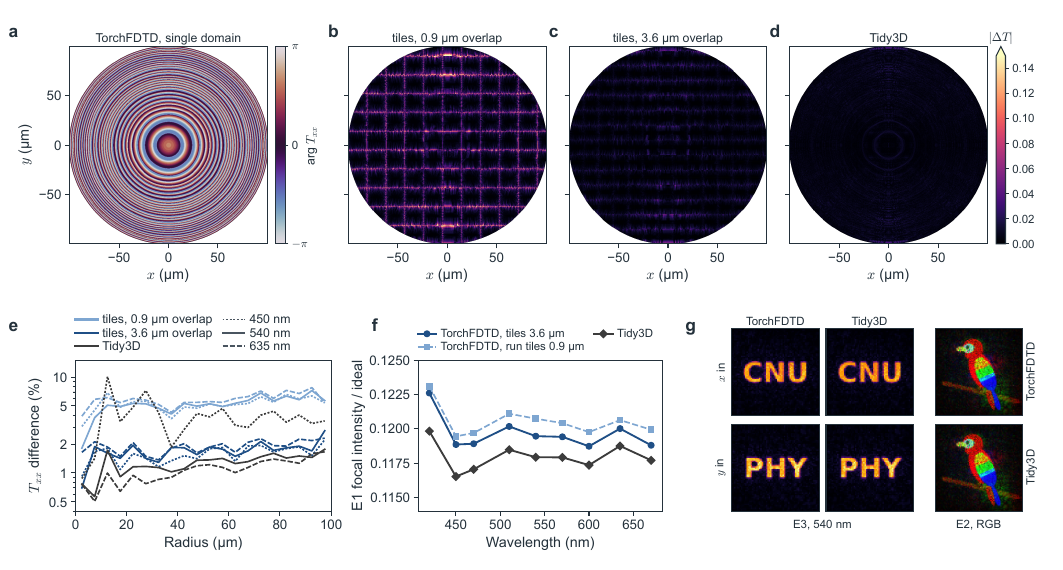}
\caption{\textbf{Tiles reproduce a single domain, and Tidy3D reproduces the optimized designs.} (a) Phase of the exit-plane map $T_{xx}$ of the meta-atom lens at \SI{540}{\nano\metre} from one TorchFDTD domain over the whole aperture. (b--d) Magnitude of the difference from (a), after one global complex factor and relative to the maximum of $|T_{xx}|$, for tiles with \SI{0.9}{\micro\metre} (b) and \SI{3.6}{\micro\metre} (c) overlaps and for Tidy3D (d), on one color scale. (e) Relative $L_{2}$ difference of $T_{xx}$ from the single domain in each \SI{5}{\micro\metre}-wide annulus, after one complex factor per wavelength as in (b--d), at 450 (dotted), 540 (solid) and \SI{635}{\nano\metre} (dashed). Averaged over the whole aperture and the three wavelengths, it is 5.8\% and 1.9\% for the two tilings and 2.2\% for Tidy3D. (f) Focal intensity of the optimized lens relative to the ideal lens at each wavelength, mean of $x$ and $y$ input, from TorchFDTD tiles with \SI{3.6}{\micro\metre} overlaps (solid), from the TorchFDTD tiles of the optimization with \SI{0.9}{\micro\metre} overlaps (dashed) and from Tidy3D. (g) Images of the optimized holograms from the TorchFDTD tiles of the optimization and from Tidy3D, on one color scale per design.}
\label{fig:accuracy}
\end{figure}

We first solve the three optimized designs as single TorchFDTD domains, which reproduce the tiled objectives to within 0.01\% (\supptab{singledomain}). On the \SI{200}{\micro\metre} meta-atom lens, which we also solved with Tidy3D at 450, 540 and \SI{635}{\nano\metre}, tiles with the \SI{3.6}{\micro\metre} overlap used for scoring differ from one TorchFDTD domain by 1.9\% in the complex near field, between 1 and 3\% from the center to the rim. That domain and Tidy3D differ by 2.2\% (Fig.~\ref{fig:accuracy}a--e). At the focus, the tiles agree with the single domain to 0.02\% with \SI{3.6}{\micro\metre} overlaps and 0.26\% with the \SI{0.9}{\micro\metre} overlaps of the optimization (\supptab{threeway}), and the single domain agrees with Tidy3D to 1.4\%. All four solutions, the two tilings, the single domain and Tidy3D, place the on-axis peaks at the same positions (\suppfig{threeway} and \suppnote{xsolver}). Tiling changes the cost of a solution, not its answer.

An independent solver reproduces the scores of the three designs. We solve the final binary layouts over the whole aperture with Tidy3D\cite{ncs_hughes2021} on their permittivity arrays and score them with the same evaluator: the designs reach 0.985 to 0.986 of their TorchFDTD objectives from tiles with \SI{3.6}{\micro\metre} overlaps (\supptab{freeformval}). Their co-polarized exit-plane maps correlate to 0.997 or better, and their focal-plane intensities and hologram images differ by 1.5 to 2.7\% in relative $L_{2}$ norm (Fig.~\ref{fig:accuracy}f,g and \suppnote{xsolver}).

\subsection*{Cost and memory scale with the aperture area}
\begin{figure}[tbp]
\centering
\includegraphics[width=\linewidth]{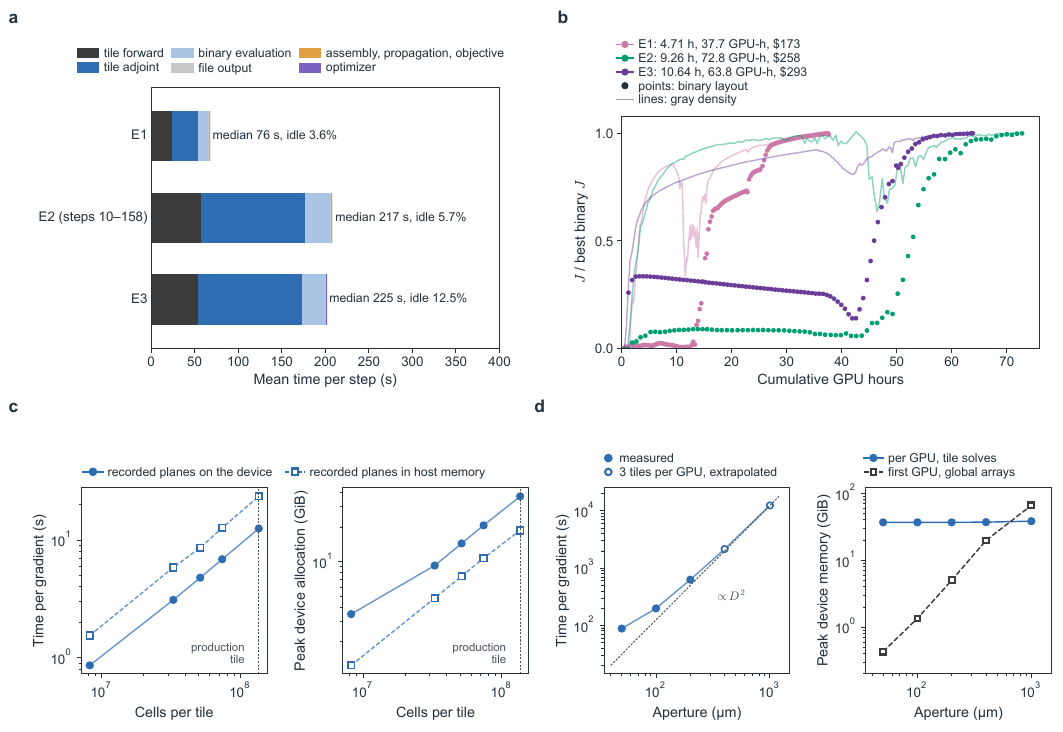}
\caption{\textbf{Tile solutions set the cost of a gradient while the tile memory per GPU stays constant.} (a) One optimization step: mean time by stage for E1, E2 (steps 10 to 158) and E3, with the median step time and the mean fraction of GPU time without recorded work. (b) Convergence against cost: objective of the binary layout (points) and of the gray density (lines), divided by the best binary objective of each run, against cumulative GPU hours, with the compute time, GPU hours and cost of each design at the node prices given in Methods. (c) One tile gradient on one H200: time and peak device allocation against the number of cells, 1,200 steps and three wavelengths, with the recorded planes on the device or in host memory; the dashed line marks the production tile. (d) Aperture scaling on four H100 GPUs: time of one full-aperture gradient with three wavelengths against the aperture diameter, measured at 50, 100 and \SI{200}{\micro\metre} (filled) and extrapolated at 400 and \SI{1000}{\micro\metre} from three solved tiles per GPU (open), with a $D^{2}$ line; and peak device memory of the tile solves per GPU and of the first GPU, which holds the global \SI{10}{\nano\metre} design grid and the pupil arrays.}
\label{fig:scaling}
\end{figure}

Tile solutions set the time of an optimization step: forward and adjoint tile solutions fill almost all of a step, and assembly, propagation, objective and optimizer together take less than 1~s (Fig.~\ref{fig:scaling}a and \suppnote{cost}). A complete design takes 4.7 to 10.6~h and \$173 to \$293 on one node, counting the wall-clock time of the optimization steps at the hourly price of the node (Fig.~\ref{fig:scaling}b and Methods). A direct FDTD calculation of the lens through its focal volume would update about 5,100 times as many cells per pass as one tiled gradient. It would need $1.6\times10^{5}$ GPU hours for 250 steps and 10~TB for its fields, against 37.7 GPU hours for the tiled optimization (\suppnote{cpu}), a decisive margin: a lens that a direct calculation would keep on an eight-GPU node for about two years takes under five hours to design.

One tile gradient can also trade time for memory: with the recorded planes in host memory, the production tile takes 23.5~s and 18.7~GiB instead of 12.5~s and 37.0~GiB (Fig.~\ref{fig:scaling}c). Tiling changes how memory grows with the aperture. A single-domain adjoint would need about 1.0~TB for the D4 quarter of the \SI{200}{\micro\metre} lens, 91\% of the memory of an eight-GPU node, and about 2.3~TB for the asymmetric \SI{150}{\micro\metre} hologram, more than one node (\suppfig{landscape}b and Methods). In contrast, a tile needs the same memory at any aperture. On four GPUs, one three-wavelength gradient takes 627~s at \SI{200}{\micro\metre}, and an extrapolation from three solved tiles per GPU gives 3.5~h for the 2{,}277 tiles of a \SI{1}{\milli\metre} aperture, with the tile solves at 36.7 to 38.2~GiB per GPU throughout (Fig.~\ref{fig:scaling}d). The tiled gradient grows as $D^{2}\log D$ in computation and as $D^{2}$ in memory, against $D^{4}$ for a direct time-domain adjoint through the focal volume, whose height and duration scale with the focal length (\supptab{scaling}).

\section*{Discussion}
Full-wave inverse design of large metasurfaces has been limited by the memory of the reverse sweep as much as by the speed of the field updates. By reconstructing the interior from the final state and from the two planes that bound the absorbers, we remove the time dimension from the volume memory. As a result, a production tile gradient of 1,200 steps fits in 37~GiB instead of 3.9~TB. Tiles that exchange no data during a solve then fix the memory of the tile solutions per GPU, and the gradient of a \SI{1}{\milli\metre} aperture extrapolates to 3.5~h on four GPUs at unchanged tile memory. The size of a design thereby becomes a question of GPU hours rather than of memory. These results establish that full-wave inverse design of aperiodic metasurfaces hundreds of wavelengths across is a computation of hours on one GPU node: the \SI{200}{\micro\metre}, nine-wavelength lens completes in 4.7~h on eight GPUs. To our knowledge, they are the first full-wave three-dimensional freeform designs of this size optimized as one layer without locally periodic unit cells or periodically repeated zones.

Our comparison with meta-atom designs shows where full-wave design matters most: where one meta-atom must serve several functions, or where the target phase steps sharply between neighbors. These are the regimes of the most demanding flat optics, broadband imaging\cite{ncs_froch2025}, full-color display\cite{ncs_so2023,ncs_gopakumar2024} and polarization multiplexing\cite{ncs_xiong2023} among them. Meta-atom libraries thereby become a baseline for large metasurfaces rather than their design model. Multifunctional aperiodic layers, where libraries fall short by up to a factor of 1.65, become open to gradient-based design at full aperture. The chain also accepts several layers or a height-resolved density, since the FDTD tiles are already three-dimensional, and takes fabrication filters and minimum-feature constraints as further differentiable stages before rasterization.

None of these ingredients is specific to optics. Reverse reconstruction needs only an explicit update that runs backwards in a lossless interior and the recorded values on two bounding planes. Overlapping tiles with an exact transpose turn an adjoint too large for one device into many independent solves, and fused stencil kernels attack the memory-bandwidth bound shared by explicit solvers on GPUs\cite{ncs_zhang2023ebisu}. The same construction applies to explicit wave solvers with a reversible update and absorbing boundaries.

\section*{Methods}

\subsection*{Electromagnetic model}
Each tile is a three-dimensional Yee grid\cite{cpc_yee1966} with a \SI{20}{\nano\metre} step in all directions and 125 cells along $z$, of which the silicon nitride layer occupies \SI{700}{\nano\metre} above a silica substrate and below air. Refractive indices are constant, 2.0 for silicon nitride and 1.444 for silica, and the permittivity of each Yee cell is obtained from the material density by the subpixel averaging described below. The time step is 0.99 times the three-dimensional Courant limit, and each tile runs for 1,200 steps. The lateral faces of a tile are periodic, and convolutional perfectly matched layers\cite{cpc_roden2000} of 12 cells terminate $z$. A soft broadband plane source in the substrate, \SI{0.55}{\micro\metre} below the layer, injects a normally incident pulse with a spectrum from 400 to \SI{700}{\nano\metre}, polarized along $x$ or $y$. A plane monitor \SI{0.5}{\micro\metre} above the layer accumulates the discrete Fourier transforms of the six field components at 420, 450, 470, 510, 540, 570, 600, 635 and \SI{670}{\nano\metre} for the lens, at 450, 540 and \SI{635}{\nano\metre} for the color hologram. The polarization-switched hologram uses the same stack and a single monitor wavelength, \SI{540}{\nano\metre}. Fields are stored in single precision and all exterior operations use double precision.

\subsection*{Discrete adjoint and recovery of forward fields}
Let $u^{n}$ denote the field and absorber state after step $n$ and $u^{n+1}=F(u^{n},\epsilon)$ the update with permittivity $\epsilon$. For an objective that depends on the recorded spectra, the reverse recursion $\bar u^{n}=(\partial F/\partial u)^{\mathsf T}\bar u^{n+1}+\partial J/\partial u^{n}$ and the accumulation $\bar\epsilon=\sum_{n}(\partial F/\partial\epsilon)^{\mathsf T}\bar u^{n+1}$ give the gradient with respect to every permittivity value in one sweep, and every update and its transpose run as fused CUDA kernels of TorchFDTD\cite{torchfdtdsoftware} (\suppnote{adjoint} and \suppnote{kernels}). The interior of a tile is lossless and nondispersive, so the forward fields are reconstructed backwards from the final state with the tangential fields of two recorded planes as boundary values, while the adjoint propagates through the absorbing layers over the whole domain; no checkpoint is stored and no forward step is replayed (\suppnote{adjoint} and \supptab{memschemes}). The single-domain references use a forward-only solver that splits the domain into slabs, one per GPU (\suppnote{adjoint}).

\subsection*{Memory of the reverse sweep}
A stored-history adjoint keeps the six single-precision field components of every cell at every step, $24N_{t}$ bytes per cell, or 3.9~TB for a production tile of $1.35\times10^{8}$ cells and 1,200 steps. The reconstruction keeps a volume state of 159 bytes per cell, independent of $N_{t}$ and of the number of recorded wavelengths, and the two recorded planes, whose size is proportional to the tile area times $N_{t}$. Computed by the solver alone on one H200 GPU, the gradient of a production tile of $1040\times1040\times125$ cells and 1,200 steps takes 12.5~s and peaks at 37.0~GiB with three or nine wavelengths, about 294 bytes per cell, or 23.5~s and 18.7~GiB with the recorded planes in host memory. A single-domain adjoint, measured at the production mesh on domains of 8 to \SI{24}{\micro\metre} with 1,200 steps, needs about 315 bytes per cell, and its memory grows linearly with the number of cells and of time steps. Extrapolated in this way, the D4 quarter of the \SI{200}{\micro\metre} lens needs about 1.0~TB and the full \SI{150}{\micro\metre} color hologram about 2.3~TB (\suppfig{landscape}b). For the fixed layer of the tiled model, the gradient memory is the tile volume plus its area times the number of steps, and its total over the aperture grows with $D^{2}$.

\subsection*{Overlapping tiles and their transpose}
Let $\rho$ be the material density on the global \SI{10}{\nano\metre} raster and $S_{j}$ the restriction of $\rho$ to the window of tile $j$, which covers its core and an overlap $o$ on every side. The tile solves give exit-plane spectra $u_{j}(S_{j}\rho)$, and the global exit plane is $u=\sum_{j}R_{j}C_{j}u_{j}$, where $C_{j}$ keeps the core of tile $j$ and $R_{j}$ places it on the global plane. The derivative of an objective $J(u)$ is
\begin{equation}
\frac{\partial J}{\partial\rho}=\sum_{j}S_{j}^{\mathsf T}\left(\frac{\partial u_{j}}{\partial\rho_{j}}\right)^{\mathsf T}C_{j}^{\mathsf T}R_{j}^{\mathsf T}\,\frac{\partial J}{\partial u},
\label{eq:tiles}
\end{equation}
in which $(\partial u_{j}/\partial\rho_{j})^{\mathsf T}$ is the tile adjoint and $S_{j}^{\mathsf T}$ adds the gradient of the whole window, overlap included, to the global raster. A density cell in an overlap therefore receives contributions from every tile whose window contains it (\suppnote{tiling}). The aperture is divided into square cores of \SI{19}{\micro\metre}, or \SI{25}{\micro\metre} for the polarization-switched hologram, and during optimization each tile extends $o=\SI{0.9}{\micro\metre}$ beyond its core on every side, so the windows of the lens are \SI{20.8}{\micro\metre} wide. The periodic lateral faces are only the boundary of each tile computation. The freeform pattern differs from window to window, the margins keep each core away from these faces, and the gradient is taken with respect to the density of the whole aperture. Each tile then computes $(20.8/19)^{2}=1.20$ times the area of its core for the lens and the color hologram and $(26.8/25)^{2}=1.15$ times for the polarization-switched hologram. The final evaluation uses \SI{15}{\micro\metre} cores with $o=\SI{3.6}{\micro\metre}$. For the D4-symmetric lens, 16 representative tiles remain, solved for $x$ and $y$ polarization as 32 tile solutions of $1040\times1040\times125$ cells and 1,200 steps per pass on the eight GPUs of one node, and the fields of the other tiles follow from the Jones matrices of the representatives under the rotations and reflections of the aperture. The color hologram solves the 64 tiles of its square aperture for $x$ polarization, and the polarization-switched hologram solves its 16 tiles for both polarizations, 32 solutions, to obtain the full Jones matrix of every tile at \SI{540}{\nano\metre}.

\subsection*{Exterior propagation, normalization and calibration}
The assembled exit plane is sampled on a pupil grid of \SI{0.25}{\micro\metre}, each pupil value being the average of $2\times2$ field samples at \SI{0.125}{\micro\metre}, and the $800\times800$ pupil of the \SI{200}{\micro\metre} aperture is zero-padded to twice its size. Each transverse component is propagated by the angular spectrum\cite{cpc_goodman2005,cpc_matsushima2009}
\begin{equation}
u(x,y,z+d)=\mathcal{F}^{-1}\left[\mathcal{F}\{u(x,y,z)\}\,\exp\left(\mathrm{i}d\sqrt{k^{2}-k_{x}^{2}-k_{y}^{2}}\right)\right],
\label{eq:asm}
\end{equation}
with transverse wavenumbers beyond $0.65k$ removed, and the longitudinal electric and transverse magnetic components follow from the transversality of each plane wave, so the Poynting flux through any observation region is available. The transpose of Eq.~\eqref{eq:asm} is the same operation with the conjugate kernel, applied to the cotangent. For the lens this replaces the \SI{333}{\micro\metre} air path of a direct calculation by two transforms per wavelength and plane. Tile spectra are divided by the incident field of a matching substrate-only reference, whose Poynting flux normalizes every wavelength, so efficiencies are fractions of the incident power. The plane monitors of TorchFDTD use the temporal kernel $\exp(+\mathrm{i}\omega t)$, and the sign of the propagation phase was checked in homogeneous air with fields recorded at $z=0.85$ and \SI{0.95}{\micro\metre}, which advance by $+k\Delta z$ to within 0.0038~rad at all nine wavelengths and both polarizations (\suppnote{calibration} and \suppfig{calibration}).

\subsection*{Distribution over GPUs}
Tile solutions within one pass are distributed over the GPUs of a node, one process per GPU, with the NCCL backend of PyTorch. The unit of work is one tile and one input polarization, and the units are split in a fixed canonical order into contiguous shares, with the smallest share on rank 0, which holds the design, the assembly, the propagation and the optimizer. Rank 0 broadcasts the design, each rank solves its units, and rank 0 gathers the exit planes in canonical order by point-to-point transfers. After the objective is evaluated, rank 0 broadcasts the exit-plane cotangent, each rank runs the adjoints of its units, and rank 0 receives the unit gradients in canonical order and adds them through the transposed window maps. The result is therefore independent of the number of GPUs, and one and two ranks give bitwise identical results for the non-symmetric path. In the optimization runs, each rank kept the recorded planes in pinned host memory for all 32 tile solutions of the lens, 16 of the 64 of the color hologram (12 in its first ten steps) and 6 of the 32 of the polarization-switched hologram, and repeated the forward solution of the other tiles in the adjoint pass. The peak allocation per rank was 45.1, 40.1 and 64.1~GiB in the three runs, the last with the $1340\times1340\times125$ tiles of the polarization-switched hologram.

\subsection*{Design parameterization and optimization}
The latent variables $\theta\in[0,1]$ live on a polar grid of \SI{10}{\nano\metre} cells in one eighth of the aperture for the D4-symmetric lens and on a Cartesian \SI{10}{\nano\metre} grid of $20000\times20000$ cells, restricted to the aperture, for the non-symmetric designs. The lens has $7.85\times10^{7}$ variables on $10{,}001\times7{,}854$ polar cells, rasterized onto $9.82\times10^{6}$ cells of the \SI{20}{\nano\metre} grid, and the color and polarization-switched holograms have $2.25\times10^{8}$ and $1.00\times10^{8}$ variables. A normalized Gaussian filter with standard deviation \SI{20}{\nano\metre} and a projection $\hat\rho=[\tanh(\beta\eta)+\tanh(\beta(\tilde\rho-\eta))]/[\tanh(\beta\eta)+\tanh(\beta(1-\eta))]$ with $\eta=0.5$ follow\cite{cpc_bourdin2001,cpc_wang2011}, and the sharpness $\beta$ starts at 2 and is raised adaptively, following the trend of the binary objective, up to 256. The projected density is rasterized onto the \SI{10}{\nano\metre} global grid, cut to the aperture, extruded through the layer and averaged onto the \SI{20}{\nano\metre} Yee grid, with the fill of the design pixels around each electric-field component averaged anisotropically along fixed interface normals, as a diagonal subpixel permittivity\cite{cpc_farjadpour2006}, in both optimization and scoring. The optimizer is Adam with learning rate 0.05, every design starts from a gray density of 0.5 with uniform noise of amplitude 0.15 (seed 20260926), and the thresholded binary layout $\hat\rho>0.5$ is evaluated separately at intervals and at the end, the final evaluation with a tile overlap of \SI{3.6}{\micro\metre}. Reported performance always refers to the binary layout, and the final layouts are shown in \suppnote{layouts} and Supplementary Figs.~\ref{fig:topview_e1}--\ref{fig:topview_e3}. The lens was optimized for 250 steps, and $\beta$ reached 256 at step 178. At step 249 the gray density reached $J_{1}=0.1214$ and its thresholded layout 0.1200. The best binary layout, 0.1205, occurred at step 248 and fills 0.442 of the aperture. The final gray density has a gray fraction of 0.064, and the best binary layout lies 0.7\% below the final gray objective. The color hologram ran for 159 steps while $\beta$ rose to 256. Its best binary layout, at step 158, has an objective of 0.676 in the optimization against 0.679 for the gray density of that step, which has a gray fraction of 0.068, and fills 0.398 of the aperture. The polarization-switched hologram ran for 190 steps with $\beta$ rising to 256, and at step 189, the last step, its binary layout reaches 0.848 against 0.851 for the gray density.
\subsection*{Objectives}
All objectives act on the $z$ component $S_{z}$ of the Poynting vector of the propagated field, normalized so that a unit plane wave carries $S_{z}=1$. For the lens, $S(\lambda,p)$ is $S_{z}$ at the focus $(0,0,f)$ for input polarization $p$ divided by the same quantity for an ideal lossless lens of the same aperture on the same grid, with pupil transmission $\exp[-\mathrm{i}k(\sqrt{r^{2}+f^{2}}-f)]$. With $\bar S_{\lambda}$ its mean over the two polarizations, the objective is the power mean $J_{1}=\bigl[\frac{1}{9}\sum_{\lambda}\bar S_{\lambda}^{\,q}\bigr]^{1/q}$ with $q=-8$, which lies between $\min_{\lambda}\bar S_{\lambda}$ and $9^{1/8}=1.32$ times that minimum. The lens is reported by the hard minimum and the mean of $\bar S_{\lambda}$ over wavelength. Its focusing efficiency is the power within 1.5 Airy radii, $1.5\times0.61\lambda/\mathrm{NA}$ with $\mathrm{NA}=0.287$, divided by the power incident on the aperture, and for comparison with the literature also the power inside a disk whose diameter is three times the focal full width at half maximum\cite{ncs_zda2025,ncs_chen2018}. The Strehl ratio is the peak focal $S_{z}$ divided by that of an ideal lens transmitting the same power as the design\cite{ncs_zda2025}. The focal full width is given in units of $\lambda/2\mathrm{NA}$, the focal shift is the axial position of the peak of $S_{z}$ relative to $f$, compared with $\lambda/\mathrm{NA}^{2}$. For the color hologram, $P_{c}$ is the power at wavelength $\lambda_{c}$ in each \SI{1.25}{\micro\metre} pixel of the \SI{120}{\micro\metre} window at $z=\SI{300}{\micro\metre}$ behind the square aperture, integrated with $4\times4$ samples per pixel and divided by the incident power, and $\eta_{c}$ its sum. The similarity $\hat S(P,t)=[\cos(P_{+},t)-\cos(1,t)]/[1-\cos(1,t)]$ is one only for an image proportional to the target $t$ and zero for a uniform window. Each wavelength contributes $J_{c}=\eta_{c}\bigl[\hat S(P_{c},t_{c})-w\sum_{c'\neq c}\max\bigl(0,\hat S(P_{c},t_{c'})-\hat S(t_{c},t_{c'})\bigr)\bigr]$ with $w=0.5$, and $J_{2}=-\tau\log\bigl[\frac{1}{3}\sum_{c}\exp(-J_{c}/\tau)\bigr]$ with $\tau=0.02$. For the polarization-switched hologram, $P_{p}$ is the total $S_{z}$ for input polarization $p$ in each \SI{1.25}{\micro\metre} pixel of the \SI{80}{\micro\metre} window at $z=\SI{200}{\micro\metre}$ behind the \SI{100}{\micro\metre} square aperture, with the same sampling and normalization, and $\eta_{p}$ its sum. The images of $x$ and $y$ input are the sums of the intensities of the two transmitted components with no analyzer. Each input contributes $J_{p}=\eta_{p}\bigl[\hat S(P_{p},t_{p})-w\max\bigl(0,\hat S(P_{p},t_{p^{\prime}})-\hat S(t_{p},t_{p^{\prime}})\bigr)\bigr]$ with $w=0.5$, where $p^{\prime}$ is the other polarization, and $J_{3}=-\tau\log\bigl[\frac{1}{2}\sum_{p}\exp(-J_{p}/\tau)\bigr]$ with $\tau=0.02$. The letter pixels of a word are the pixels of its target above half of the maximum, 332 for CNU and 306 for PHY. The gradients of the lens and color-hologram objectives with respect to the exit-plane field agree with central differences along random directions to $4\times10^{-10}$ and $3\times10^{-9}$ in double precision, the second on the full \SI{150}{\micro\metre} pupil, and the gradient of the polarization-switched hologram objective agrees to $4\times10^{-9}$ on its full \SI{100}{\micro\metre} pupil with both cross-talk terms active (\suppnote{objectives}).

\subsection*{Meta-atom designs}
The library contains square silicon nitride posts \SI{700}{\nano\metre} tall on a \SI{280}{\nano\metre} square lattice, with widths from 60 to \SI{200}{\nano\metre} in \SI{10}{\nano\metre} steps and an empty cell, which keeps a minimum width of \SI{60}{\nano\metre} and a minimum gap of \SI{80}{\nano\metre}. The lattice period is below $\lambda/n_{\mathrm{SiO_2}}$ at \SI{420}{\nano\metre}, so only the zeroth order propagates. The zeroth-order transmission $t(w,\lambda)$ of each width for light incident from the substrate was computed with TORCWA\cite{ncs_kim2023} in double precision on a \SI{1}{\nano\metre} raster with $13\times13$ Fourier orders, and $11\times11$ orders serve as a convergence check. The two truncations differ by at most 0.018 in $t$ and 0.020~rad in phase, the energy balance $|t|^{2}+|r|^{2}=1$ holds to $1.6\times10^{-12}$, and every width transmits at least 0.65 of the incident power. Lattice sites lie at $\pm(140+280k)$~nm, posts are placed only where their largest footprint lies inside the aperture, and each post is aligned so that both edges fall on the \SI{10}{\nano\metre} design grid, which keeps D4 symmetry and the gap rule. For the lens, each site takes the width that minimizes $\sum_{\lambda}|t(w,\lambda)-\exp(\mathrm{i}\phi_{\lambda})|^{2}$ over the nine wavelengths, with $\phi_{\lambda}(r)=-(2\pi/\lambda)(\sqrt{r^{2}+f^{2}}-f)+C_{\lambda}$. This nine-wavelength match sets the group delay of each site together with its phase\cite{ncs_shrestha2018,ncs_chen2018}, and the nine offsets $C_{\lambda}$ are chosen for the whole aperture by coordinate descent from eight starts. For the color hologram, a Gerchberg--Saxton iteration\cite{ncs_gerchberg1972} of 100 steps per channel, with the same angular-spectrum propagation, unit-modulus phasors on the post lattice and the target power imposed in each \SI{1.25}{\micro\metre} pixel of the window, gives phases $\phi_{c}$ that place 97 to 98\% of the power in the window, and each site takes the width minimizing $\sum_{c}|t(w,\lambda_{c})-\exp[\mathrm{i}(\phi_{c}+C_{c})]|^{2}$ with three global offsets. The color-hologram baseline used in the main text couples the iteration to the library. Starting from these phases, it inserts the realized library transmissions into the pupil at every iteration, reselects the widths jointly for the three wavelengths and resets the offsets, and it keeps the best of 400 iterations. The hologram posts fill the square aperture on a $536\times536$ lattice. For the polarization-switched hologram, the library holds rectangular posts with side lengths from 60 to \SI{200}{\nano\metre} in \SI{10}{\nano\metre} steps and the empty cell, 225 shapes whose $2\times2$ Jones matrices were computed in the same way at \SI{540}{\nano\metre}. An axis-aligned post transmits $x$ and $y$ light independently as $\mathrm{diag}(t_{xx},t_{yy})$ with cross terms below $3\times10^{-11}$, so each site is assigned the retarder $\mathrm{diag}[\exp(\mathrm{i}\phi_{x}),\exp(\mathrm{i}\phi_{y})]$ with the phases of a Gerchberg--Saxton iteration of 100 steps for each word, and takes the post that minimizes the squared complex error of both transmissions with one global offset per polarization. As for the color hologram, the baseline of the main text then iterates the loop 400 times with the realized transmissions, reselects the posts jointly and keeps the best iterate. The posts fill the \SI{100}{\micro\metre} aperture on a $356\times356$ lattice and are rasterized onto the \SI{10}{\nano\metre} design grid. The baselines are library designs, scored by the same tiled evaluator with \SI{3.6}{\micro\metre} overlaps, propagation and objective as the freeform designs. The phase step of a meta-atom layout is $(\Delta_{x}\phi^{2}+\Delta_{y}\phi^{2})^{1/2}$, with $\Delta_{x}\phi$ and $\Delta_{y}\phi$ the wrapped phase differences of the selected meta-atom transmissions to the neighboring sites along $x$ and $y$ on the \SI{280}{\nano\metre} lattice, maximized over the wavelengths or input polarizations, and Extended Data Table~\ref{tab:gains} gives its 95th percentile over the lattice sites. The target lens phase alone gives 1.17~rad per period, so for the lens the step comes mostly from the discrete library selection.

\subsection*{Accuracy tests}
Directional derivatives were tested with central differences of step $\delta=5\times10^{-3}$ in density along gradient-aligned and random unit directions (\supptab{gradverify} and \supptab{objectivefd}). The full-domain comparisons solve the same layer as one domain with the same mesh, duration, source, boundaries and monitors, and report relative $L_{2}$ norms $\|a-b\|/\|b\|$ of the complex exit-plane fields, of the propagated intensity and of the density gradient over the design region, the relative objective difference and the cosine similarity of the gradients, all evaluated in double precision from the saved arrays (\suppnote{overlap}, \supptab{overlap} and \suppfig{overlap}). The single-domain reference of the \SI{200}{\micro\metre} meta-atom lens was solved by TorchFDTD with the D4 quarter divided into slabs over four GPUs and by Tidy3D with the same quarter symmetry, on the permittivity array of the production tiles at 450, 540 and \SI{635}{\nano\metre} (\suppnote{validation} and \supptab{threeway}). The three optimized designs were solved in the same way as single domains, the lens as its D4 quarter, with 1,200 steps and scored by the same evaluator (\supptab{singledomain}). The independent-solver comparison re-scores the final binary layouts over the whole aperture with Tidy3D, with matched materials, source spectrum, mesh and exit-plane coordinates (\suppfig{e3tidy}), after a small periodic patch solved by every code (\supptab{xsmall}). It used Tidy3D 2.10.2 with the per-component permittivity of the production raster supplied on its own Yee grid without subpixel averaging, a uniform \SI{20}{\nano\metre} mesh, periodic lateral faces, the D4 quarter for the lens, perfectly matched layers along $z$, a plane-wave pulse polarized along $x$ or $y$, a run time of 60~fs (1,575 steps) and a field-decay shutoff of $10^{-5}$. The runs cost 1.6 FlexCredits for the lens and for each polarization of the polarization-switched hologram and 3.5 for the color hologram. The tile gradient was compared with that of FDTDX 0.6.2, an independent JAX-based GPU FDTD solver, on windows of the polarization-switched hologram on one H200 GPU (\suppnote{xsolver}). Layer thickness and edge bias are evaluated in \suppnote{robust}.

\subsection*{Cost accounting}
Every timed quantity is a wall-clock time between device synchronizations. The time of a design iteration includes tile forward and adjoint solutions, assembly, propagation, objective, communication and the optimizer, and excludes one-time construction and file output, which are reported separately. Device memory is the peak Torch allocation, host memory the peak resident set size. For the $1040\times1040\times125$ tile of the lens and the color hologram, the gradient computed by the solver alone peaks at 37.0~GiB on one H200, the forward and reverse call inside the optimization driver at 39.7~GiB on one H200, and the tile phase of the multi-GPU aperture benchmark at 36.7 to 38.2~GiB per H100. The cost of a design is the wall-clock time of its optimization steps multiplied by the hourly price of the node, including idle ranks and binary evaluations. The CPU model was fitted to short runs of the production tile physics with 32 to 96 lateral cells, 125 vertical cells and 16 or 32 steps on four threads, with the largest run held out, and is applied to cell-update counts. The fit gives $8.7~\mathrm{ms}+3.07\times10^{-8}~\mathrm{s}\times N_{\mathrm{cell}}N_{t}$, predicts the held-out run to within 4.1\% (\suppfig{cpufit}, \supptab{cpu} and \suppnote{cpu}), and for the production tile layouts gives 101 to 202~h per iteration on four threads, or 2.9 years for the 250 steps of the lens.

\subsection*{Computing resources}
The computations used rented GPU nodes on RunPod. The lens was optimized for 250 steps in 4.71~h on eight H200 GPUs (37.7 GPU hours and \$173 at \$36.72 per hour), with a median step of 75.6~s for 32 tile solutions of $1040\times1040\times125$ cells per pass and about \SI{53}{\second} of forward and adjoint solves per step. The polarization-switched hologram was optimized for 190 steps in 10.64~h on six H200 GPUs (63.8 GPU hours and \$293 at \$27.54 per hour), with a median step of 225.3~s for 32 tile solutions of $1340\times1340\times125$ cells. The color hologram was optimized for 159 steps in 9.26~h, on six H200 GPUs for steps 0 to 9 and on eight NVIDIA H100 GPUs (\$27.92 per hour) from step 10, in total 72.8 GPU hours and \$258, with a median step of 217.5~s on the H100 node for 64 tile solutions of $1040\times1040\times125$ cells (\suppnote{cost}). The single-domain and tiled solutions of the \SI{200}{\micro\metre} meta-atom lens took about 20~min of computation and the aperture-scaling benchmark 46~min, both on four H100 GPUs, and the full-wave scoring of the final and meta-atom layouts with \SI{3.6}{\micro\metre} overlaps took 23~min on eight H100 GPUs. The single-domain solutions of the three optimized designs took about 25~min of node time on eight H100 GPUs (\$11.8). The comparison with FDTDX took 32~min on one H200 GPU and that with Meep 31~min, with Meep on 20 CPU cores, and the other single-GPU benchmarks and gradient tests took 1.8~h on one H200 SXM GPU. The Tidy3D solutions used 14.8 FlexCredits on the Flexcompute cloud.

\subsection*{Use of generative AI}
Generative AI tools (OpenAI GPT-6 Astra and Anthropic Claude Fable 5.1 and Opus 5.5) assisted with code implementation, test generation and benchmark scripting for TorchFDTD and the design, analysis and figure scripts of this work. All code was checked by the automated test suite and by the numerical validation reported here, and the authors reviewed the code and take full responsibility for the content of the publication.

\subsection*{Data availability}
The source data behind every figure and table, the final layouts of the three designs and of the meta-atom baselines, the meta-atom libraries and the optimization histories are available in the TorchFDTD repository at \url{https://github.com/hyoseokp/TorchFDTD} (folder \path{examples/full-aperture-tiled-adjoint}) and archived on Zenodo as version 1.1.6 at \url{https://doi.org/10.5281/zenodo.23189322} (all versions: \url{https://doi.org/10.5281/zenodo.22928834}). The transmitted exit-plane fields and all scores can be regenerated from the released layouts with the evaluation script.

\subsection*{Code availability}
TorchFDTD is available at \url{https://github.com/hyoseokp/TorchFDTD} under the MIT license and archived on Zenodo as version 1.1.6 at \url{https://doi.org/10.5281/zenodo.23189322} (all versions: \url{https://doi.org/10.5281/zenodo.22928834}; ref.~\citen{torchfdtdsoftware}). The design drivers, objectives, targets, configurations and evaluation scripts of the three designs are included as examples in the repository and in the Zenodo archive.

\subsection*{Acknowledgements}
This work was supported by the BK21 FOUR Program, the Chungnam National University Research Grant 2026, the Basic Science Research Program through the National Research Foundation of Korea (NRF) funded by the Ministry of Education (RS-2020-NR049597), the creation of the quantum information science R\&D ecosystem (based on human resources) through the NRF funded by the Ministry of Science and ICT (RS-2023-00256050), the Global--Learning \& Academic research institution for Master's, PhD students, and Postdocs (G-LAMP) Program of the NRF funded by the Ministry of Education (RS-2025-25442707), the NRF grant funded by the Korea government (Ministry of Science and ICT) (RS-2025-24533176), the Technology Innovation Program Development Program (RS-2023-00235080) funded by the Ministry of Trade, Industry \& Energy (MOTIE, Korea), and the NRF grant funded by the Korea government (MSIT) (RS-2026-25606142).
\subsection*{Author contributions}
H.P. conceived the study, developed TorchFDTD and the tiled adjoint framework, designed the objectives and meta-atom baselines, ran the optimizations, the validations against single-domain solutions and independent solvers and the cost measurements, analyzed the data, prepared the figures and wrote the manuscript. Y.P. supervised the work and edited the manuscript.
\subsection*{Competing interests}
The authors declare no competing interests.
\bibliographystyle{unsrt}
\bibliography{refs}

\clearpage
\section*{Extended Data}
\begin{table}[h]
\renewcommand{\tablename}{Extended Data Table}
\centering
\caption{\textbf{Freeform and meta-atom designs of the three examples.} E1, broadband lens; E2, color hologram; E3, polarization-switched hologram. Functions, simultaneous conditions per site; phase step, 95th percentile of the neighboring phase step of the meta-atom layout (Methods); LPA, locally periodic prediction for the meta-atom layout; meta-atom and freeform, objectives of the binary layouts from full-wave tiles with \SI{3.6}{\micro\metre} overlaps; ratio, freeform over meta-atom.}
\label{tab:gains}
\footnotesize
\setlength{\tabcolsep}{4pt}
\begin{tabular}{lllcccccc}
\toprule
 & Aperture & Symmetry & Functions & Phase step (rad) & LPA & Meta-atom & Freeform & Ratio \\
\midrule
E1 & \SI{200}{\micro\metre} disk & D4 & $9\lambda\times2$ pol. & 3.05 & 0.0776 & 0.0728 & 0.1196 & 1.64 \\
E2 & \SI{150}{\micro\metre} square & none & $3\lambda$ & 2.79 & 0.456 & 0.409 & 0.673 & 1.65 \\
E3 & \SI{100}{\micro\metre} square & none & 2 pol. & 2.08 & 0.679 & 0.680 & 0.844 & 1.24 \\
\bottomrule
\end{tabular}
\end{table}

\clearpage
\setcounter{section}{0}
\setcounter{figure}{0}
\setcounter{table}{0}
\setcounter{equation}{0}
\renewcommand{\thesection}{\arabic{section}}
\renewcommand{\figurename}{Supplementary Fig.}
\renewcommand{\tablename}{Supplementary Table}
\renewcommand{\theequation}{S\arabic{equation}}
\renewcommand{\theHsection}{S\arabic{section}}
\renewcommand{\theHfigure}{S\arabic{figure}}
\renewcommand{\theHtable}{S\arabic{table}}
\renewcommand{\theHequation}{S\arabic{equation}}
\makeatletter
\renewcommand{\section}{\@startsection{section}{1}{\z@}{3.0ex plus 1ex minus .2ex}{1.2ex}{\normalfont\large\bfseries Supplementary Note~}}
\makeatother
\singlespacing
\setlength{\parskip}{3pt}
\renewcommand{\topfraction}{0.85}
\renewcommand{\textfraction}{0.1}
\renewcommand{\floatpagefraction}{0.75}
\begin{center}
{\large\bfseries Supplementary Information for\\[4pt] Inverse design of large-scale freeform meta-optics by breaking the memory wall of full-wave simulation\par}
\vskip 1.5em
{\large Hyoseok Park and Yeonsang Park\\[2pt]\small Department of Physics, Chungnam National University, Daejeon, Republic of Korea\par}
\end{center}
\vskip 1.5em
\section{Discrete adjoint, recorded-interface reconstruction and single-domain solutions}
\label{si:adjoint}
The solver advances the electric and magnetic fields of the Yee grid in alternating half steps and updates the memory variables of the convolutional absorbing layers in the same kernels\cite{torchfdtdsoftware}. We write one full step as $u^{n+1}=F(u^{n},\epsilon)+s^{n+1}$, where $u^{n}$ holds the fields and absorber memories and $s$ is the source. Because $F$ is linear in $u$ for a fixed permittivity, its transpose is again a sequence of local stencil operations. The curl that maps magnetic to electric fields becomes, under transposition, the curl with the opposite difference direction and the same metric factors, and the scalar coefficients of the absorber recursion act on the cotangents in reverse order. The derivative with respect to the inverse permittivity of a cell is the product of the electric cotangent and the curl of the magnetic field at that cell, summed over all steps. TorchFDTD implements each of these transposed updates as a separate CUDA kernel alongside its forward kernel, and the reverse sweep returns the derivative of the discrete model. Each electric or magnetic update, together with the absorber memories, is one fused kernel, and the fused kernels of one step are captured with source injection and spectral accumulation in one CUDA graph that is replayed every step. The spectral monitors are linear in the fields, so their transposes inject the objective cotangent as a source of the reverse sweep, and no automatic-differentiation graph of the time stepping is recorded.

\subsection*{Discretization}
The six field components occupy the standard staggered Yee positions, and one step of the nonmagnetic update reads
\begin{equation}
E^{n+1}=E^{n}+S\,\epsilon^{-1}\,\mathcal{C}^{-}\!\left(H^{n+1/2}\right)+s^{n+1},\qquad
H^{n+3/2}=H^{n+1/2}-S\,\mathcal{C}^{+}\!\left(E^{n+1}\right),
\label{eq:yee}
\end{equation}
where $S=c\Delta t/h$ with the mesh step $h=\SI{20}{\nano\metre}$, $\mathcal{C}^{\pm}$ are the forward and backward difference curls, and the soft source increment $s$ is added after the electric update. The time step is $\Delta t=0.99\,h/(c\sqrt{3})$, 0.99 times the three-dimensional Courant limit. Fields start from zero, the first electric sample lies at $\Delta t$ and the first magnetic sample at $3\Delta t/2$, and every discrete Fourier transform keeps this half-step offset. The plane monitors interpolate the six components of each Yee cell to a common point and apply the magnetic half-step phase before the signed Poynting flux is formed. Each absorbing face of $L=12$ cells along $z$ carries one memory variable $\psi$ per transverse derivative $D$, and the derivative used in the curl is $y=D/\kappa+\psi$ with $\psi\leftarrow b\psi+cD$, $b=\exp[-(\sigma/\kappa+\alpha)S]$ and $c=(b-1)\sigma/(\sigma\kappa+\alpha\kappa^{2})$. The dimensionless profiles are $\sigma=40\rho^{3}/(L+1)$, $\kappa=1$ and $\alpha=10^{-8}$, where $\rho=d/(L+1)$ is the depth $d$ of the derivative position into the face in cells. The transpose of the recursion is $\bar\psi_{\mathrm{mem}}=\bar\psi+\bar y$, $\bar D=\kappa^{-1}\bar y+c\,\bar\psi_{\mathrm{mem}}$ and $\bar\psi_{\mathrm{old}}=b\,\bar\psi_{\mathrm{mem}}$.

\subsection*{Recorded-interface reconstruction}
The reverse sweep needs the forward fields in decreasing order of $n$. The interior of a tile between the two absorbing layers is lossless and nondispersive, and the permittivity differs from a fixed background only inside a reconstruction interval $\Omega_{R}$ of the $z$ layers $a$ to $b$. During the forward pass the solver records at every step the tangential electric field $(E_x,E_y)$ in layer $b+1$ above $\Omega_{R}$ and the tangential magnetic field $(H_x,H_y)$ in layer $a-1$ below it, and it keeps the state of $\Omega_{R}$ at the final step. These two layers are the only values outside $\Omega_{R}$ that the curls inside $\Omega_{R}$ read, so stepping Eq.~\eqref{eq:yee} backwards inside $\Omega_{R}$ recovers every earlier interior field (Algorithm below). The absorbing layers are never inverted and their forward fields are never needed. The adjoint is linear in the fields, so it propagates through the absorbers over the whole domain with the transposed absorber recursion, and only the material cotangent, which is confined to $\Omega_{R}$, reads the reconstructed forward fields. No checkpoint is stored and no forward step is replayed.

\medskip
\noindent\fbox{\parbox{0.96\linewidth}{\small
\textbf{Algorithm.} Time-domain adjoint with recorded interfaces and interior reconstruction.\\[2pt]
\textbf{Input:} permittivity $\epsilon$ that differs from a fixed background only inside $\Omega_{R}$ (layers $a\ldots b$), $N_{t}$ steps.\quad \textbf{Output:} objective $J$ and cotangent $\bar\epsilon^{-1}$ on $\Omega_{R}$.
\begin{algorithmic}[1]
\State $E,H,\psi\gets0$ \Comment{forward pass}
\For{$n=0,\ldots,N_{t}-1$}
  \State record $T_H^{n}\gets(H_x,H_y)^{n+1/2}$ in layer $a-1$
  \State $E^{n+1}\gets E^{n}+S\epsilon^{-1}\mathcal{C}^{-}(H^{n+1/2})+s^{n+1}$ \Comment{absorber memories updated inside $\mathcal{C}^{\pm}$}
  \State record $T_E^{n}\gets(E_x,E_y)^{n+1}$ in layer $b+1$
  \State $H^{n+3/2}\gets H^{n+1/2}-S\,\mathcal{C}^{+}(E^{n+1})$
  \State accumulate the monitor spectra
\EndFor
\State keep $E^{N_t},H^{N_t+1/2}$ on $\Omega_{R}$; evaluate $J$ and the seeds $\partial J/\partial u^{n}$
\State $\bar u\gets0$, $\bar\epsilon^{-1}\gets0$ \Comment{reverse pass}
\For{$n=N_{t}-1,\ldots,0$}
  \State restore $(E_x,E_y)^{n+1}$ in layer $b+1$ from $T_E^{n}$
  \State $H^{n+1/2}\gets H^{n+3/2}+S\,\mathcal{C}^{+}(E^{n+1})$ on $\Omega_{R}$
  \State restore $(H_x,H_y)^{n+1/2}$ in layer $a-1$ from $T_H^{n}$
  \State $\bar u\gets(\partial F/\partial u)^{\mathsf T}\bar u+\partial J/\partial u^{n}$ on the whole domain, absorbers included
  \State $q\gets\mathcal{C}^{-}(H^{n+1/2})$ on $\Omega_{R}$
  \State $\bar\epsilon^{-1}\gets\bar\epsilon^{-1}+S\,\bar E^{n+1}\odot q$ \Comment{fused with the next line}
  \State $E^{n}\gets E^{n+1}-S\epsilon^{-1}q-s^{n+1}$ on $\Omega_{R}$
\EndFor
\end{algorithmic}
\textit{Retained forward data:} $4N_{t}N_{\perp}$ values for the two recorded layers and $6N_{R}$ values for the terminal state, with $N_{\perp}=N_{x}N_{y}$ lateral cells and $N_{R}$ cells in $\Omega_{R}$.}}
\medskip

The record holds four transverse components per step, 16 bytes per lateral cell per step in single precision. A production tile of $1040\times1040$ lateral cells and 1,200 steps therefore records $16\times1200\times1040^{2}$ bytes, or 20.8~GB (19.3~GiB), and the terminal state of its 35 design layers holds 0.91~GB. Supplementary Table~\ref{tab:memschemes} compares this record with the other ways of supplying the forward fields. The gradient of the production tile peaks at 37.0~GiB with the planes on the device and at 18.7~GiB with the planes in host memory, where the device keeps two blocks of 32 steps (1.1~GB); the difference, 18.3~GiB, is the device share of the record.

The solver measures the round-off of the single-precision reconstruction at the end of every reverse sweep, where the reconstructed state must return to the zero initial state. On one H200 GPU, after 1,200 steps of the production physics the residual is $5.7\times10^{-7}$ of the forward field norm in relative $L_{2}$ norm and $4.9\times10^{-7}$ in peak for the production tile, and 5.66 to $5.73\times10^{-7}$ for windows of $256^{2}$ to $1040^{2}$ lateral cells, identical with the planes on the device or in host memory (Supplementary Table~\ref{tab:gradverify}). On a small CPU fixture of $6\times7\times20$ cells with a $z$-normal plane source, in single precision, the recorded-interface gradient agrees with that of the checkpoint-replay adjoint, which repeats the forward arithmetic exactly, to $1.6\times10^{-7}$ in relative $L_{2}$ norm.

\begin{table}[tbp]
\centering
\small
\caption{Ways of supplying the forward fields to the reverse sweep, evaluated for the production tile of $1040\times1040\times125$ cells, 1,200 steps and 35 design layers in single precision. The device and host columns give the retained forward data. The binomial schedule\cite{cpc_griewank2000} with $C=8$ states of 3.66~GB each, fields and absorber memories, recomputes no step more than five times. The measured peak allocations of the whole gradient call on one H200, which include fields, absorber memories, material maps, adjoint fields and spectral buffers, are 37.0~GiB for the restricted interval with device planes and 18.7~GiB with host planes.}
\label{tab:memschemes}
\footnotesize
\setlength{\tabcolsep}{3.5pt}
\begin{tabular}{lllll}
\toprule
Scheme & Device & Host & Extra forward work & Gradient \\
\midrule
Full forward history, whole tile & 3.9~TB & none & none & exact \\
Electric history of the design layers & 545~GB & none & none & exact \\
Binomial checkpoints, $C=8$ & 29~GB or none & none or 29~GB & $\le5$ per step & exact \\
Recorded interfaces, 99 lossless layers & 23.3~GB & none & none & round-off \\
Recorded interfaces, 35 design layers & 21.7~GB & none & none & round-off \\
As above, planes in host memory & 2.0~GB & 20.8~GB & none & round-off \\
\bottomrule
\end{tabular}
\end{table}

\subsection*{Multi-GPU single domains}
The single-domain references are computed by a forward-only solver that splits the domain along $x$ into contiguous slabs, one per GPU, each holding the full $y$ and $z$ extent. Before every half step, neighboring slabs exchange one $x$ row of the two tangential components through NCCL point-to-point transfers: the magnetic row moves to the right neighbor before the electric update, and the electric row moves to the left neighbor before the magnetic update. Every update repeats the elementwise arithmetic of the TorchFDTD reference step in the same order, so the result does not depend on the number of GPUs. The solver uses the production mesh, materials, source, 1,200 steps and exit-plane monitor. Absorbing layers of 12 cells terminate $z$ and the open lateral faces, which lie \SI{2}{\micro\metre} beyond the aperture, and the D4 quarter of a lens is bounded on its two lower faces by electric and magnetic mirror walls. The meta-atom lens was solved as its D4 quarter on four GPUs, and the three optimized designs on eight GPUs, the lens as its D4 quarter and the two holograms over their whole apertures. On a $20\times\SI{20}{\micro\metre}$ window of the broadband lens ($1000\times1000\times125$ cells, \SI{20}{\nano\metre}, 1,200 steps, 450, 540 and \SI{635}{\nano\metre}), the fused resident solver, the TorchFDTD reference step and the slab solver with one and with two slabs on one H200 SXM GPU give bitwise identical exit-plane spectra of $E_x$, $E_y$, $H_x$ and $H_y$.

\section{The tiled model and its derivative}
\label{si:tiling}
\begin{figure}[tbp]
\centering
\includegraphics[width=\linewidth]{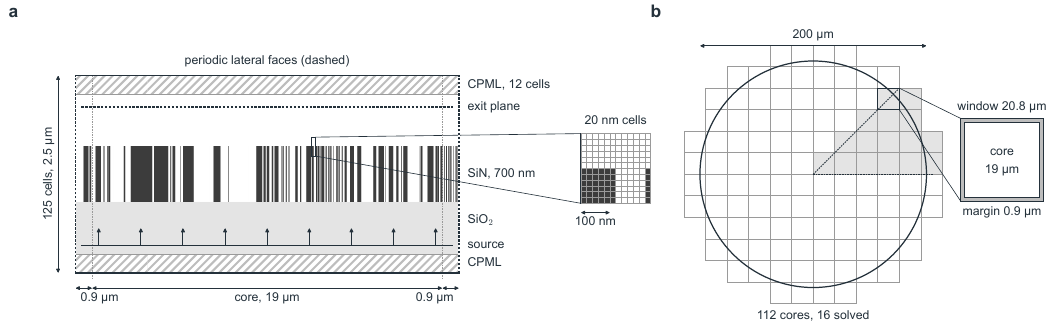}
\caption{\textbf{Tile geometry.} (a) Section of one production tile of the lens at the physical heights, with its absorbing layers, source and exit plane, one \SI{10}{\nano\metre} row of the optimized layout as the silicon nitride layer and the \SI{20}{\nano\metre} Yee cells at one pillar edge; the vertical scale is stretched. (b) The 112 tile cores of \SI{19}{\micro\metre} over the \SI{200}{\micro\metre} lens aperture, with the 16 representatives solved under D4 symmetry shaded and one \SI{20.8}{\micro\metre} window enlarged to scale.}
\label{fig:tilegeom}
\end{figure}

The assembled exit plane $u=\sum_{j}R_{j}C_{j}u_{j}(S_{j}\rho)$ and its derivative, given in Methods of the main text, define the tiled model. The restriction $S_{j}$ copies the density of the window of tile $j$, $u_{j}$ is the exit-plane spectrum of the tile, $C_{j}$ keeps its core and $R_{j}$ places the core on the global plane. $R_{j}^{\mathsf T}$ reads the global cotangent at the core of tile $j$, $C_{j}^{\mathsf T}$ pads it with zeros over the overlap, the tile adjoint maps it to a density cotangent over the whole window, and $S_{j}^{\mathsf T}$ adds that window to the global raster. The overlap of the window therefore receives gradient from every tile that contains it, which is required for the derivative of the assembled model. Supplementary Fig.~\ref{fig:tilegeom} shows the geometry of one production tile and the tile cores of the \SI{200}{\micro\metre} lens.

Supplementary Table~\ref{tab:gradverify} lists the gradient tests, including the directional derivative tests of the complete production chain on a \SI{12}{\micro\metre} aperture with the \SI{20}{\nano\metre} mesh, \SI{10}{\nano\metre} latent grid, \SI{0.25}{\micro\metre} pupil and \SI{0.125}{\micro\metre} field samples of the production geometry, isotropic material averaging and the calibrated propagation sign. The chain includes the polar latent grid, filter, projection, aperture raster, material averaging on the Yee faces, FDTD, Jones-matrix assembly of both polarizations and vector propagation. The three test objectives are a focal spot, four spots and a ring pattern. Supplementary Table~\ref{tab:objectivefd} lists the same test for a single \SI{6}{\micro\metre} domain with six recorded field components, nine wavelengths and each input polarization. For the non-symmetric path, the window map between the global \SI{10}{\nano\metre} raster and the tiles passes a dot-product transpose test on three tiles to $5.1\times10^{-9}$ to $4.3\times10^{-8}$, and a directional central difference of the complete chain on a \SI{6}{\micro\metre} fixture of nine tiles with a test objective agrees with the adjoint to $3.9\times10^{-5}$ on one H200 SXM GPU. With the launcher of Methods, one and two ranks give bitwise identical designs, gradients and objectives for this path.

\begin{table}[tbp]
\centering
\small
\caption{\textbf{Gradient verification.} CD, central difference; on and off, the reverse-sweep kernel policy; aligned, gradient-aligned direction. The FDTDX rows give the relative $L_{2}$ difference of the full gradient of one tile with 1,200 steps and three wavelengths on one H200, with the cosine similarity in parentheses (Supplementary Fig.~\ref{fig:fdtdxadj}). The other rows give the relative error $|g_{\mathrm{adj}}-g_{\mathrm{FD}}|/|g_{\mathrm{FD}}|$ between the adjoint directional derivative and a central difference. On the \SI{12}{\micro\metre} aperture of the production chain the error is $7.4\times10^{-8}$ to $2.8\times10^{-6}$ along gradient-aligned directions and $4.5\times10^{-5}$ to $1.5\times10^{-4}$ along random directions; the kernel policy is the reverse-sweep memory-traffic policy of Supplementary Note~\ref{si:kernels}, and each case took about 21~s on one H200 SXM GPU. The \SI{6}{\micro\metre} cases are listed in Supplementary Table~\ref{tab:objectivefd} and the \SI{30}{\micro\metre} cases in Supplementary Note~\ref{si:overlap}.}
\label{tab:gradverify}
\footnotesize
\setlength{\tabcolsep}{4pt}
\begin{tabular}{llll}
\toprule
Comparison & Case & Reference & Rel. difference \\
\midrule
Tile gradient & $256^{2}\times125$ cells & FDTDX\cite{cpc_mahlau2024} & 2.4\% (0.99970) \\
 & $512^{2}\times125$ cells & FDTDX & 3.1\% (0.99952) \\
 & $768^{2}\times125$ cells & FDTDX & 3.9\% (0.99925) \\
Matched discretization & $64^{3}$ box, 128 steps & FDTDX, checkpointed & $7.1\times10^{-8}$ \\
 & $64^{3}$ box, 128 steps & FDTDX, reversible & $1.0\times10^{-7}$ \\
\addlinespace
Chain, \SI{12}{\micro\metre} & Focal spot, off, aligned & CD & $7.42\times10^{-8}$ \\
 & Focal spot, off, random & CD & $1.47\times10^{-4}$ \\
 & Four spots, off, aligned & CD & $1.50\times10^{-6}$ \\
 & Four spots, off, random & CD & $4.47\times10^{-5}$ \\
 & Ring pattern, off, aligned & CD & $2.48\times10^{-6}$ \\
 & Ring pattern, off, random & CD & $7.61\times10^{-5}$ \\
 & Focal spot, on, aligned & CD & $2.80\times10^{-6}$ \\
 & Focal spot, repeat, off, aligned & CD & $1.95\times10^{-6}$ \\
\addlinespace
Optical objectives, 3 & \SI{6}{\micro\metre}, 2 polarizations & CD & $0.4$--$4.6\times10^{-6}$ \\
Tiles, \SI{50}{\nano\metre} mesh & \SI{30}{\micro\metre}, full domain & CD & $5.4\times10^{-5}$ \\
 & \SI{30}{\micro\metre}, nine tiles & CD & $5.5\times10^{-5}$ \\
\addlinespace
Reconstruction, H200 & $256^{2}$--$1040^{2}\times125$, 1,200 steps & zero initial state & $5.7\times10^{-7}$ \\
Recorded interfaces, CPU & $6\times7\times20$ cells & checkpoint replay & $1.6\times10^{-7}$ \\
\bottomrule
\end{tabular}
\end{table}

\begin{table}[tbp]
\centering
\small
\caption{Directional derivative tests on a single \SI{6}{\micro\metre} domain with a \SI{0.5}{\micro\metre} margin, \SI{20}{\nano\metre} mesh, nine wavelengths and the calibrated propagation sign. Time to gradient includes the forward pass, the recorded forward pass and the adjoint, median of three runs; the first case of each run includes kernel compilation. Peak memory is the peak Torch allocation on one H200 SXM GPU.}
\label{tab:objectivefd}
\begin{tabular}{lllll}
\toprule
Objective & Input & Relative error & Time to gradient (s) & Peak memory (GB) \\
\midrule
Focal spot & $E_x$ & $2.58\times10^{-6}$ & 3.80 & 4.26 \\
Four spots & $E_x$ & $1.81\times10^{-6}$ & 2.14 & 4.29 \\
Ring pattern & $E_x$ & $4.04\times10^{-7}$ & 2.14 & 4.29 \\
Focal spot & $E_y$ & $4.64\times10^{-6}$ & 2.14 & 4.29 \\
Four spots & $E_y$ & $2.00\times10^{-6}$ & 2.14 & 4.29 \\
Ring pattern & $E_y$ & $1.13\times10^{-6}$ & 2.14 & 4.29 \\
\bottomrule
\end{tabular}
\end{table}

\section{Propagation phase and flux calibration}
\label{si:calibration}
\begin{figure}[tbp]
\centering
\includegraphics{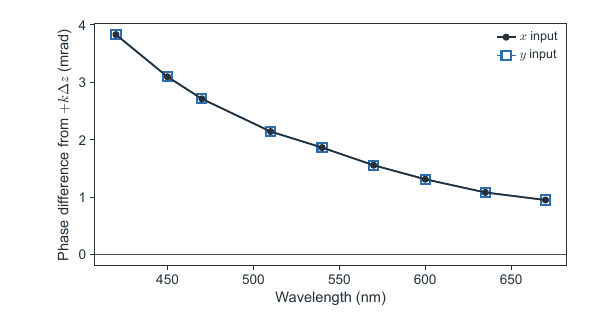}
\caption{Propagation phase in homogeneous air. Difference between the measured phase advance over \SI{0.1}{\micro\metre} and $+k\Delta z$ at the nine design wavelengths for both input polarizations.}
\label{fig:calibration}
\end{figure}

The plane monitors of TorchFDTD accumulate spectra with the temporal kernel $\exp(+\mathrm{i}\omega t)$, the complex conjugate of the kernel used by its point monitors. With this convention a wave traveling toward positive $z$ acquires the phase $+k\Delta z$, and the angular spectrum of Methods carries the positive sign. We checked the sign on a periodic homogeneous-air domain with the production mesh of \SI{20}{\nano\metre} and 1,200 steps, recording the six field components on planes at $z=0.85$ and \SI{0.95}{\micro\metre}. For both input polarizations and all nine wavelengths the phase advance between the planes matches $+k\Delta z$ to 0.0038~rad (Supplementary Fig.~\ref{fig:calibration}). A second homogeneous calculation in silica gives the incident Poynting flux of each wavelength, which normalizes all efficiencies.

\section{Overlap convergence with absorbing lateral boundaries}
\label{si:overlap}
\begin{figure}[tbp]
\centering
\includegraphics{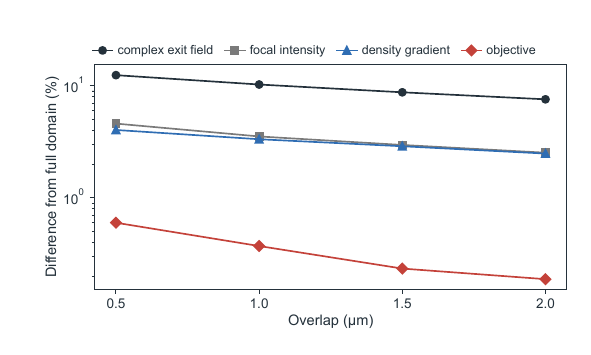}
\caption{Differences of the tiled model from the full domain against the lateral overlap, for the \SI{30}{\micro\metre} test of Supplementary Table~\ref{tab:overlap}.}
\label{fig:overlap}
\end{figure}

\begin{figure}[tbp]
\centering
\includegraphics{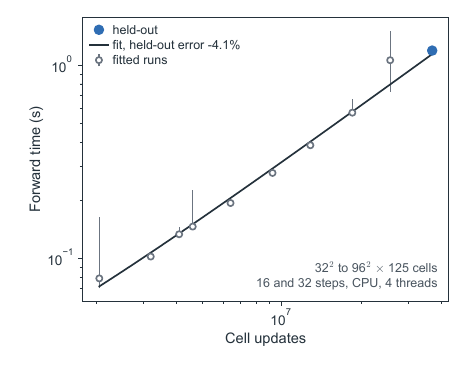}
\caption{CPU cost model of the production tile physics. Forward time of short CPU runs on four threads against the number of cell updates, for $32^{2}$ to $96^{2}\times125$ cells and 16 and 32 steps: medians of three repetitions with their range (open markers), the nonnegative fit (line) and the held-out run (filled marker).}
\label{fig:cpufit}
\end{figure}

The overlap test of the main text uses a \SI{30}{\micro\metre} square gray density between 0.15 and 0.85, generated from uniform random values on a \SI{0.15}{\micro\metre} grid with a fixed seed and interpolated bilinearly to the \SI{50}{\nano\metre} mesh. The \SI{31}{\micro\metre} domain is \SI{3}{\micro\metre} tall with absorbing layers on all faces, the source sheet extends through the lateral absorbers, and the \SI{700}{\nano\metre} layer has permittivity $1+3\rho$ on silica. Each model runs 600 steps with two host checkpoints and records the exit plane at nine wavelengths from 420 to \SI{670}{\nano\metre} for $E_x$ input. The objective is the propagated intensity \SI{47.4}{\micro\metre} beyond the exit plane, the focal distance of an NA~0.3 lens of that width, weighted by a Gaussian detector of standard deviation \SI{1.5}{\micro\metre}. The design region holds $3.6\times10^{5}$ density values. The adjoint agrees with a central difference to $5.4\times10^{-5}$ for the full domain and $5.5\times10^{-5}$ for the tiles with \SI{1.5}{\micro\metre} overlap. Supplementary Table~\ref{tab:overlap} lists all differences, and Supplementary Fig.~\ref{fig:overlap} plots them against the overlap. Widening the overlap from 0.5 to \SI{2.0}{\micro\metre} raises the cosine similarity between the tiled and full-domain gradients from 0.9992 to 0.9997 and brings the relative $L_2$ difference of the complex exit-plane field from 12.4\% to 7.6\%. The objective converges faster, agreeing with the full domain to 0.60\% and then 0.19\%, because propagation to the observation plane spreads the near-field difference at the seams over the whole aperture.

\begin{table}[tbp]
\centering
\small
\caption{Nine tiles with \SI{10}{\micro\metre} cores against the full \SI{30}{\micro\metre} domain on a \SI{50}{\nano\metre} mesh. Relative $L_2$ differences of the complex exit-plane field, the propagated intensity and the density gradient, relative difference of the objective and cosine similarity of the gradients. Time to gradient, median of three runs, and peak Torch allocation on one H200 SXM GPU, tiles solved one after another. The full domain takes 153.4~s and 5.85~GB.}
\label{tab:overlap}
\setlength{\tabcolsep}{4pt}
\begin{tabular}{rrrrrrrr}
\toprule
Overlap & Field & Intensity & Objective & Gradient & Cosine & Time & Peak \\
(\micron) & (\%) & (\%) & (\%) & (\%) & & (s) & (GB) \\
\midrule
0.5 & 12.44 & 4.60 & 0.602 & 4.04 & 0.99920 & 164.6 & 2.90 \\
1.0 & 10.27 & 3.54 & 0.373 & 3.34 & 0.99946 & 188.9 & 3.21 \\
1.5 & 8.75 & 2.97 & 0.235 & 2.89 & 0.99960 & 211.7 & 3.53 \\
2.0 & 7.58 & 2.53 & 0.189 & 2.49 & 0.99971 & 233.7 & 3.87 \\
\bottomrule
\end{tabular}
\end{table}

The field error falls with distance from the nearest seam. In bands of \SI{0.25}{\micro\metre}, the root-mean-square relative field difference decreases from 22.2\% next to the seam to 10.0\% at 1.25 to \SI{1.5}{\micro\metre} from it for an overlap of \SI{0.5}{\micro\metre}, and from 12.0\% to 6.4\% for an overlap of \SI{2.0}{\micro\metre}.

\section{Memory traffic of the reverse sweep}
\label{si:kernels}
The FDTD update is limited by memory bandwidth, so the reverse sweep, which reconstructs forward fields and applies the transposed updates, reads and writes more bytes per cell than the forward pass. In summed kernel time on one H200, the Yee updates take 4.0~ms of the 4.2~ms forward step of a production tile, and the transposed updates and the reconstruction take 3.7 and 2.0~ms of its reverse step (Fig.~2b of the main text). A kernel policy for the reverse sweep reduces this traffic. It combines four measures. The reconstruction interval $\Omega_{R}$ of Supplementary Note~\ref{si:adjoint} is restricted to the 35 design layers, so the terminal state and the reverse reconstruction cover 35 instead of 99 lossless layers, while the recorded layers move to the faces of the smaller interval and keep their size. The inverse electric update and the accumulation of the material cotangent run in one fused kernel, so the curl of the magnetic field is evaluated once per reverse step. With the recorded layers in host memory, blocks of 32 steps pass through two pinned host buffers on a separate copy stream, so the transfer of one block overlaps the computation of the next and the device holds two blocks. A forward call that cannot request a gradient, such as a binary evaluation, records no layers and no terminal state. Supplementary Table~\ref{tab:kernels} compares the policy with the default on a tile of $980\times980\times125$ cells, close in size to the production tiles, with scalar material, device records and 512 steps. The saved fields are identical under both policies and the gradients differ by $3.97\times10^{-7}$ in relative $L_2$ norm. On an H200 GPU, Nsight Systems kernel times on a production tile of $1040\times1040\times125$ cells give effective bandwidths, the minimum byte count per cell divided by the kernel time per cell, of 61\% and 56\% of 4.8~TB/s for the forward updates, 66\% and 61\% for the adjoint updates and 42\% for the electric recovery.

On the tile of Supplementary Table~\ref{tab:kernels}, $980\times980\times125=1.2\times10^{8}$ cells, the forward pass advances 512 steps in 2.46~s, or $2.50\times10^{10}$ cell updates per second, and the reverse-sweep policy lowers the forward and reverse call from 7.83 to 5.88~s and the reverse sweep from 5.21 to 3.34~s.

Supplementary Table~\ref{tab:fused} compares the fused kernels with an unfused execution of the same discrete update on one H200 SXM GPU, for the production tile of $1040\times1040\times125$ cells with diagonal permittivity, three wavelengths and 512 steps. The unfused path is the PyTorch tensor-operation path of TorchFDTD, which evaluates every curl, absorber update, source injection and field update as a separate kernel: 152 kernels per forward step and 333 per step of the forward and reverse call, against 11 and 29 for the fused path and 23 per step of the forward and reverse call with the reverse-sweep policy, the count shown in Fig.~2a of the main text. Both paths reconstruct the whole lossless interior. Their recorded spectra are bitwise identical, and their gradients differ by $2.7\times10^{-7}$ in relative $L_2$ norm with a cosine similarity of $1-2\times10^{-13}$. The fused forward step takes 4.37~ms, or $3.09\times10^{10}$ cell updates per second, against 37.0~ms unfused, a factor of 8.5. The reverse step takes 10.5 against 85.3~ms, a factor of 8.2, and the forward and reverse call 10.9 against 63.8~s with a peak allocation of 25.9 against 32.3~GiB. The minimum traffic of a step is 84 bytes per cell for the forward updates, and for the reverse step 84 bytes per cell for the adjoint updates plus 144 bytes per reconstructed cell for the magnetic and electric recovery and the material cotangent, or 120 bytes with the reverse-sweep policy, which recovers the electric field and accumulates the cotangent in one kernel. Divided by the step times, these counts give effective bandwidths of 54\% and 53\% of 4.8~TB/s for the fused forward and reverse steps and 6.4\% and 6.5\% for the unfused steps. With the reverse-sweep policy, the reverse step takes 5.93~ms at 56\% of the peak and the call 5.43~s, with a gradient that differs from the fused one without the policy by $2.9\times10^{-7}$.

On a $24\times\SI{24}{\micro\metre}$ domain of the production mesh, the fused update kernels and their transposes advance 25.7 billion cells per second on one H200, 35 to 39 times the rate of Meep\cite{ncs_oskooi2010} on 20 CPU cores for the same layer.

\begin{table}[tbp]
\centering
\small
\caption{One production tile ($1.2\times10^{8}$ cells, 512 steps) on one H200 SXM GPU with and without the reverse-sweep policy. Forward is a pass without recording. The forward and reverse call consists of a recorded forward pass and the reverse sweep. Forward and call are medians, and the recorded forward and the reverse sweep the fastest, of three runs after one warm-up.}
\label{tab:kernels}
\begin{tabular}{lrrrr}
\toprule
Policy & Forward & Forward and & Recorded & Reverse \\
 & (s) & reverse (s) & forward (s) & sweep (s) \\
\midrule
Default & 2.51 & 7.83 & 2.61 & 5.21 \\
Reverse-sweep policy & 2.46 & 5.88 & 2.52 & 3.34 \\
\bottomrule
\end{tabular}
\end{table}

\begin{table}[tbp]
\centering
\small
\caption{Fused and unfused field updates on one H200 SXM GPU for a production tile of $1040\times1040\times125$ cells with diagonal permittivity, three wavelengths and 512 steps. Unfused, the same discrete update executed as separate PyTorch tensor operations; policy, the reverse-sweep policy of Supplementary Table~\ref{tab:kernels}. Step times are synchronized loop times per step without setup, and the call time is the wall time of the forward and reverse call. Bandwidth is the minimum byte count of a step divided by its time, in percent of 4.8~TB/s. Peak is the peak Torch allocation of the call. Medians of three repetitions after one warm-up.}
\label{tab:fused}
\footnotesize
\setlength{\tabcolsep}{4pt}
\begin{tabular}{lrrrrrrr}
\toprule
Path & Forward & Rate & Bandwidth & Reverse & Bandwidth & Call & Peak \\
 & step (ms) & (Gcell/s) & (\%) & step (ms) & (\%) & (s) & (GiB) \\
\midrule
Unfused & 37.0 & 3.65 & 6.4 & 85.3 & 6.5 & 63.8 & 32.3 \\
Fused & 4.37 & 30.9 & 54.2 & 10.5 & 53.3 & 10.9 & 25.9 \\
Fused, policy & 4.24 & 31.9 & 55.7 & 5.93 & 55.9 & 5.43 & 25.9 \\
\bottomrule
\end{tabular}
\end{table}

\section{CPU timing model and memory bounds}
\label{si:cpu}
The CPU model uses the native CPU reference path of TorchFDTD with the production physics: \SI{20}{\nano\metre} mesh, 125 vertical cells, the silicon nitride layer on silica, periodic lateral faces, 12 absorbing cells along $z$, single precision, the 400 to \SI{700}{\nano\metre} source, nine monitored wavelengths and a plane monitor. Lateral sizes of 32 to 96 cells were run for 16 and 32 steps on four threads, each point with one warm-up and three timed repetitions (Supplementary Table~\ref{tab:cpu}). A nonnegative fit to the relative errors of nine medians, with the largest point held out, gives the forward time
\begin{equation}
T=8.69\times10^{-3}~\mathrm{s}+3.072\times10^{-8}~\mathrm{s}\times N_{\mathrm{cell}}N_{t}.
\end{equation}
It predicts the held-out forward median of 1.190~s as 1.141~s, a difference of $-4.13$\%, and the held-out construction-and-forward time to $-3.28$\%.

At the production mesh a square \SI{200}{\micro\metre} domain of 125 vertical cells holds $10^{4}\times10^{4}\times125=1.25\times10^{10}$ cells. Its six single-precision field components occupy 279.40~GiB, and with two scalar permittivity maps 372.53~GiB, before absorber memories, observation buffers and adjoint storage. The production tile layouts give the following estimates. The production runs use 32 tile solutions of $1040\times1040\times125$ cells per pass for the lens (16 tiles, two polarizations), 64 of $1040\times1040\times125$ cells for the color hologram and 32 of $1340\times1340\times125$ cells for the polarization-switched hologram (16 tiles, two polarizations), each for 1,200 steps. The model gives 4,984~s for one forward tile solution of $1040^{2}\times125$ cells and 8,274~s for $1340^{2}\times125$ cells on four threads, so one forward pass takes 44.3~h for the lens, 88.6~h for the color hologram and 73.5~h for the polarization-switched hologram. The reverse pass is taken as 1.28 times the forward pass, the ratio of tile adjoint to tile forward time per step in the lens run, where the recorded planes were kept and no forward solution was repeated. One iteration then takes an estimated 101~h for the lens, 202~h for the color hologram and 168~h for the polarization-switched hologram, and the 250 steps of the lens and the 190 steps of the polarization-switched hologram take $2.5\times10^{4}$ and $3.2\times10^{4}$~h, or 2.9 and 3.6 years. On the GPUs the same 250 lens steps used 37.7 GPU hours. A direct FDTD calculation of the lens through its focal volume is estimated in the same way. A D4 quarter of the aperture holds $5000\times5000$ lateral cells without lateral absorbers, and the domain extends from the 125 cells of the tile to the focal plane, $125+16{,}667=16{,}792$ cells along $z$, so it holds $4.2\times10^{11}$ cells. With $c\Delta t=\SI{11.43}{\nano\metre}$, the marginal ray needs 30,443 steps to travel $(R^{2}+f^{2})^{1/2}=\SI{348.0}{\micro\metre}$, which with the 1,200 steps of the tiles gives 31,643 steps and $1.33\times10^{16}$ cell updates per forward pass and polarization. At the forward rate of 25.7~Gcell/s measured for TorchFDTD on one H200 on a $24\times\SI{24}{\micro\metre}$ domain of the production mesh, this is 143 GPU hours per forward pass, 655 GPU hours for the forward and reverse passes of both polarizations, and $1.6\times10^{5}$ GPU hours, or \$0.75 million at \$4.59 per GPU hour, for 250 steps. One forward pass of both polarizations updates $2.66\times10^{16}$ cells, about 5,100 times the $5.19\times10^{12}$ cell updates of the 32 tile solutions of one tiled forward pass, and the 250 steps need about 4,300 times the 37.7 GPU hours of the whole tiled optimization. The six single-precision field components of this domain alone occupy 10~TB, nine times the device memory of an eight-GPU H200 node.

\begin{table}[tbp]
\centering
\small
\caption{Median CPU forward times of the production physics on four threads. Each median is taken over three timed repetitions after one warm-up.}
\label{tab:cpu}
\begin{tabular}{rrrr}
\toprule
Lateral cells & Cells & 16 steps (s) & 32 steps (s) \\
\midrule
32 & 128{,}000 & 0.0787 & 0.1332 \\
40 & 200{,}000 & 0.1021 & 0.1934 \\
48 & 288{,}000 & 0.1460 & 0.2768 \\
80 & 800{,}000 & 0.3850 & 1.0611 \\
96 & 1{,}152{,}000 & 0.5679 & 1.1903 (held out) \\
\bottomrule
\end{tabular}
\end{table}

\section{Objectives and targets}
\label{si:objectives}
All three objectives are evaluated on the propagated fields of the angular spectrum of Methods with efficiencies normalized by the incident flux of the substrate-only reference.

\emph{Broadband lens.} The focus lies at $f=\SI{333.333}{\micro\metre}$, so $R/f=0.30$, and the marginal ray reaches it at $\sin\theta_{m}=R/(R^{2}+f^{2})^{1/2}=0.287$. The optical path difference between the marginal and the axial ray is $(R^{2}+f^{2})^{1/2}-f=\SI{14.68}{\micro\metre}$. The objective uses the on-axis focal intensity relative to an ideal lossless lens, averaged over $x$ and $y$ input, and the power mean with $q=-8$ over the nine wavelengths. A gradient weight proportional to $\bar S^{\,q-1}$ gives a wavelength 10\% below the others 2.6 times their weight at any overall scale of $\bar S$. The efficiency disk has radius $1.5\times0.61\lambda/0.287$, from 1.34 to \SI{2.13}{\micro\metre} across the band, and an ideal lens puts 0.89 of the incident power inside it. For the paraxial Airy pattern, a disk out to the first zero holds 0.84 of the focal power and a disk of three full widths in diameter 0.86. The power mean with $q=-8$ exceeds the minimum over the nine wavelengths by at most a factor $9^{1/8}=1.32$, so the hard minimum is reported separately.

After the 250 steps of the lens run, the binary layout of step 249 has $\bar S_{\lambda}$ of 0.1232, 0.1181, 0.1156, 0.1188, 0.1149, 0.1262, 0.1233, 0.1231 and 0.1208 at 420, 450, 470, 510, 540, 570, 600, 635 and \SI{670}{\nano\metre}, and $J_{1}=0.1200$. Scored with \SI{3.6}{\micro\metre} overlaps, the binary layout of step 248 reaches a worst-wavelength focal intensity of 0.119 of the ideal lens and a band-averaged value of 0.120. Its focusing efficiency ranges from 0.104 at \SI{450}{\nano\metre} to 0.135 at \SI{540}{\nano\metre}, the Strehl ratio relative to an ideal lens transmitting the same power from 0.144 to 0.181, the transmission of the aperture from 0.68 to 0.83 and the power inside a disk of three focal full widths in diameter from 0.099 to 0.130 of the incident power. The depth of focus $\lambda/\mathrm{NA}^{2}$ is 5.1 to \SI{8.1}{\micro\metre} across the band. Averaged over $x$ and $y$ input, the focal intensity of the binary layout of step 248 is lowest at \SI{600}{\nano\metre}. Its focal peak lies within \SI{0.91}{\micro\metre}, or 0.13 depths of focus, of $f$ at every design wavelength except \SI{540}{\nano\metre}, where it lies \SI{2.02}{\micro\metre}, or 0.31 depths of focus, closer to the lens, and the focal full width at half maximum is 1.01 to 1.10 times $\lambda/2\mathrm{NA}$.

A thin-element reference treats the layer as a scalar phase element of height $h(x,y)\in[0,\SI{0.7}{\micro\metre}]$ that transmits $\exp[2\pi\mathrm{i}(n_{\lambda}-1)h/\lambda]$ with unit amplitude, with the dispersive index of silicon nitride, and maximizes $J_{1}$ by 300 Adam steps from the wrapped phase of a \SI{540}{\nano\metre} lens. It reaches 0.130, or 0.126 with a constant index of 2.0 and 0.109 from a random start. The continuous-band bound of Presutti and Monticone for waveguide-based dielectric metasurfaces\cite{ncs_presutti2020}, $\Delta\omega/\omega_{c}\le(H\Delta n/f)\,\Theta(\mathrm{NA})$ with $\Theta(x)=(1-x^{2})^{1/2}/[1-(1-x^{2})^{1/2}]$ in air, equals $H\Delta n/[(R^{2}+f^{2})^{1/2}-f]$. With $H=\SI{0.7}{\micro\metre}$, $\Delta n=1$ and $\mathrm{NA}=0.287$ it allows $\Delta\omega/\omega_{c}=0.048$, about a tenth of the 0.46 spanned by the design band. A layer of height $H$ supplies a group delay that matches the required path difference within a central zone of radius about $(2f\Delta n_{g}H)^{1/2}\approx22$ to \SI{26}{\micro\metre} for a group-index contrast $\Delta n_{g}$ of 1 to 1.5, or about 5 to 7\% of the aperture area.

\emph{Color hologram.} The target is a procedurally generated macaw-style parrot on a black ground, with a red body and head, yellow, green, cyan and blue wing bands and a brown branch, stored as linear RGB on $96\times96$ pixels of \SI{1.25}{\micro\metre} (\path{e2_parrot_rgb.npz}, SHA256 9be4f26ecb4d9cba\dots). The blue, green and red channels are assigned to 450, 540 and \SI{635}{\nano\metre}. The hologram has a \SI{150}{\micro\metre} square aperture, and the window at $z=\SI{300}{\micro\metre}$ subtends an image-side numerical aperture of 0.2425, so the coherent resolution $\lambda/(2\mathrm{NA})$ is 0.93, 1.11 and \SI{1.31}{\micro\metre} at the three wavelengths. The crosstalk term penalizes only resemblance to another channel beyond that of the target channels themselves. The projection steepness reached 256, and the run stopped at step 159, when the binary objective had gained no more than 0.3\% at $\beta=256$. Scored with \SI{3.6}{\micro\metre} overlaps at 450, 540 and \SI{635}{\nano\metre}, the binary layout of step 158 reaches similarities $\hat S$ of 0.911, 0.856 and 0.879 to its own channel, efficiencies into the window of 0.739, 0.796 and 0.766 and crosstalk excesses, summed over the other two channels, of 0.003, 0.019 and 0. The meta-atom hologram reaches similarities of 0.776, 0.739 and 0.755, efficiencies of 0.712, 0.787 and 0.688 and crosstalk excesses of 0.464, 0.181 and 0.041.

\emph{Polarization-switched hologram.} The target is the two words CNU and PHY, white letters on a black ground, stored as two images of $64\times64$ pixels of \SI{1.25}{\micro\metre} (\path{e3_cnu_phy_sq100.npz}, SHA256 079622464d6652bd\dots). The image of $x$ input is CNU and that of $y$ input is PHY, both the total $S_{z}$ at $z=\SI{200}{\micro\metre}$ in an \SI{80}{\micro\metre} window behind the \SI{100}{\micro\metre} square aperture, so the image-side numerical aperture is 0.243. The letters occupy 332 and 306 pixels at half of the maximum of the target. The objective is the one of the color hologram with the two inputs in place of the wavelengths. The cosine between a uniform window and the two targets is 0.308 and 0.296. The run covers 190 steps on six H200 GPUs (16 tiles of \SI{25}{\micro\metre} cores with \SI{0.9}{\micro\metre} margins, 32 tile solutions per pass, $10^{8}$ latent variables), with 172~s per step for the forward and reverse passes and 226~s when the binary layout is also evaluated, and 10.6~h in total. The projection steepness reached 256 at step 169, and the run ended at step 189, the step limit of its stop rule. The best binary objective of the complete run is 0.84795 at step 189, for a gray objective of 0.85083. The leak is the power of an input in the letter pixels that belong only to the other word divided by the power in its own letter pixels. For $x$ and $y$ input the binary layout of step 189 gives $J_{p}=0.8466$ and 0.8494, similarities of 0.9751 and 0.9753, a crosstalk excess of 0.0053 for $x$ input and 0 for $y$ input and leaks of 0.0588 and 0.0724. The gray layout of the same step has an objective of 0.8508 and a gray fraction of 0.066. Scored with \SI{3.6}{\micro\metre} overlaps, the binary layout has crosstalk excesses of 0.006 and 0 for $x$ and $y$ input, against 0.171 and 0.148 for the meta-atom hologram.

The lens target \path{e1_achromat_focus.npz} has SHA256 bedc3fbb\dots, and the history of the lens run SHA256 c2684b85f45436e8\dots. In double precision, the adjoint derivatives of the objectives with respect to the exit-plane field agree with central differences along random complex directions to $4.0\times10^{-10}$ for the lens on a small fixture and to $2.9\times10^{-10}$ for the color hologram on a small fixture with four active cross-talk terms. For the color hologram on its full \SI{150}{\micro\metre} square pupil of $600\times600$ samples, with the configuration of the production run, they agree to $1.6\times10^{-9}$ and $3.0\times10^{-9}$ along two random directions, to $1.3\times10^{-6}$ along a third direction on which the derivative is about 400 times smaller, and to $3.2\times10^{-8}$ for Gerchberg--Saxton inputs with the blue and red targets exchanged, which activate three cross-talk terms. At a step of $10^{-4}$ instead of $10^{-5}$ these values are $3.0\times10^{-10}$, $2.3\times10^{-9}$, $6.4\times10^{-7}$ and $3.2\times10^{-9}$. For the polarization-switched hologram on its full \SI{100}{\micro\metre} pupil they agree to $1.4\times10^{-10}$ and $6.0\times10^{-10}$ along two random directions, to $1.3\times10^{-7}$ along a third direction on which the derivative is about 300 times smaller, and to $4.0\times10^{-9}$ for an input that activates both cross-talk terms. On a small fixture the agreement is $3.0\times10^{-11}$ to $3.4\times10^{-8}$, the largest value again with both cross-talk terms active.

\section{Meta-atom libraries and designs}
\label{si:baselines}
The post library consists of square silicon nitride posts of height \SI{700}{\nano\metre} and refractive index 2.0 on silica of index 1.444, with air above, on a square lattice of \SI{280}{\nano\metre}. Light enters from the substrate at normal incidence. Widths run from 60 to \SI{200}{\nano\metre} in steps of \SI{10}{\nano\metre}, and an empty cell is also allowed. The upper width keeps a gap of \SI{80}{\nano\metre} between neighboring posts. Because $\SI{280}{\nano\metre}<\SI{420}{\nano\metre}/1.444$, only the zeroth order propagates in the substrate and in air at every design wavelength, so the zeroth-order transmission and reflection conserve the power. The complex transmission of each width at 420, 450, 470, 510, 532, 540, 570, 600, 635 and \SI{670}{\nano\metre} was computed with TORCWA 0.1.4.2\cite{ncs_kim2023} in double precision with Laurent factorization, a \SI{1}{\nano\metre} raster, so every width is represented without partial pixels, and $13\times13$ Fourier orders, with $11\times11$ orders as the convergence check. Because only the zeroth order propagates, $|t|^{2}+|r|^{2}=1$ tests the energy balance of each entry, and the square posts give no cross-polarized transmission. The record gives a maximum energy residual of $1.6\times10^{-12}$ and cross-polarized amplitudes below $1.1\times10^{-11}$. Between 11 and 13 orders the transmission changes by at most 0.018 and its phase by at most 0.020~rad, and the smallest power transmission in the width range is 0.65. The phase spanned by widths from 60 to \SI{200}{\nano\metre} is 6.82, 5.32, 5.38, 4.65, 4.32, 4.21, 3.84, 3.54, 3.25 and 3.01~rad at 420, 450, 470, 510, 532, 540, 570, 600, 635 and \SI{670}{\nano\metre}. The production library \path{library_o13.npz} has SHA256 a41a357a2dc9f803\dots\ and the check library \path{library_o11.npz} c872b0ba2375d3aa\dots. Both are included in the repository (folder \path{examples/full-aperture-tiled-adjoint/libraries}), together with the dispersive and rectangular-post libraries described below.

A dispersive library for the lens adds elements with C4v symmetry in the same layer: 15 square holes of 60 to \SI{200}{\nano\metre} in a continuous silicon nitride film and 33 crosses with arms 60 to \SI{120}{\nano\metre} wide and up to \SI{200}{\nano\metre} long, 64 elements with the posts and the empty cell. It was computed with the same solver, orders and wavelengths (SHA256 563d7d8f84cd278d\dots). Its energy residual is below $6\times10^{-12}$, its post entries agree with the post library to $1.6\times10^{-11}$, and 90\% of its entries change by at most 0.005 between 11 and 13 orders. Four holes of 110, 130, 160 and \SI{190}{\nano\metre} lie on guided-mode resonances of the film between 420 and \SI{470}{\nano\metre}, change by 0.08 to 0.16 between truncations, and are excluded. Over the band, the median spread of group delay across elements is 2.05~fs for the posts and 2.45~fs for the whole library, against about \SI{49}{\femto\second} required by the lens, while the phase coverage widens from 3.0 to 5.6~rad for the posts to 5.5 to 5.9~rad for the whole library. Neighboring elements must also respect the fabrication rules, which limit a post or cross next to a film cell to \SI{120}{\nano\metre} (\SI{160}{\nano\metre} at a corner contact) and every film strip to at least \SI{60}{\nano\metre}. Enforcing them changes 26\% of the sites, and the locally periodic prediction of $J_{1}$ is then 0.052 for the dispersive library, against 0.078 for the posts and 0.090 for the dispersive library without the rules. The post library is the lens baseline of the main text.

The selection rules for the three targets are given in Methods. The library also predicts the performance of each baseline under the locally periodic approximation. For the color hologram on its \SI{150}{\micro\metre} square aperture, the locally periodic model predicts $J_{2}=0.258$ for the baseline that matches independent channel phases after the iteration and 0.456 for the baseline that iterates with the library, which is used in the main text. The same independent phases realized by ideal lossless elements, each transmitting $\exp(\mathrm{i}\phi_{c})$ with unit amplitude at its own wavelength, give 0.894, an upper bound for any single-layer library realization of those phases. For the library-coupled baseline the predicted window efficiencies at 450, 540 and \SI{635}{\nano\metre} are 0.77, 0.80 and 0.70, the Pearson correlations 0.83, 0.80 and 0.82 and the structural similarities 0.49, 0.63 and 0.25. For the independent-phase baseline the values are 0.71, 0.70 and 0.62 (efficiency), 0.88, 0.73 and 0.66 (correlation) and 0.33, 0.33 and 0.11 (structural similarity).

The library of the polarization-switched hologram has 225 rectangular posts (SHA256 9a9eaf4c\dots) with axis-aligned edges, changes by at most 0.012 between 11 and 13 orders, conserves energy to $10^{-12}$, has off-diagonal Jones terms below $3\times10^{-11}$ and transmits at least 0.97, and its largest retardance is 1.08~rad. The baseline of the main text iterates independent Gerchberg--Saxton phases for $x$ and $y$ input 100 times, then couples the iteration to the library for 400 iterations and keeps the best iterate, the 380th, on a $356\times356$ lattice of sites with a $400\times400$ pupil. The locally periodic model predicts $J_{3}=0.6794$, with similarities of 0.9212 and 0.9372, crosstalk excesses of 0.1865 and 0.1597, leaks of 0.0972 and 0.1330 and a fill fraction of 0.274 on the \SI{10}{\nano\metre} grid. The first iteration without the library starts from 0.6077 and the library-coupled iteration reaches 0.6794. The layout and the images are shown in Fig.~4f,g of the main text. Scored with the full-wave tiles and \SI{3.6}{\micro\metre} overlaps, the meta-atom designs reach $J_{1}=0.0728$ for the lens, $J_{2}=0.409$ for the color hologram and $J_{3}=0.680$ for the polarization-switched hologram, against 0.0776, 0.456 and 0.679 predicted by the locally periodic model.

\section{Cross-solver verification}
\label{si:xsolver}
\label{si:validation}
The final binary layouts of the freeform and meta-atom designs are re-scored with Tidy3D, an independent GPU FDTD solver\cite{ncs_hughes2021}, over their whole apertures, with the same materials, source spectrum, mesh, exit-plane coordinates and propagation to the observation planes, using Tidy3D 2.10.2 with the settings given in Methods. The differences are reported below and in Supplementary Table~\ref{tab:freeformval}. At the production mesh the overlap was tested on the full \SI{200}{\micro\metre} meta-atom lens with the calibrated propagation (Supplementary Table~\ref{tab:threeway}). Averaged over 450, 540 and \SI{635}{\nano\metre}, tiles with \SI{0.9}{\micro\metre} overlaps differ from one TorchFDTD domain by 5.83\% in the exit-plane field and 0.26\% in the focal-plane intensity, and tiles with \SI{3.6}{\micro\metre} overlaps by 1.86\% and 0.02\%.

The Jones maps $T$ that enter the objectives are compared with independent solvers in three steps. A small periodic patch is solved with every code on the same permittivity array. The final layouts are then re-solved over their full apertures with Tidy3D. Finally the tiled production evaluations of the \SI{200}{\micro\metre} meta-atom lens and of the three optimized designs are compared with single-domain multi-GPU solutions by TorchFDTD itself.

\subsection*{Small periodic patch}
The patch is a lattice-aligned \SI{4}{\micro\metre} square window of a final layout with periodic lateral faces, taken at $x,y=10$ to \SI{14}{\micro\metre} for the polarization-switched hologram and at $x=60$ to \SI{64}{\micro\metre}, $y=0$ to \SI{4}{\micro\metre} for the broadband lens. The wavelength is \SI{540}{\nano\metre}, the \SI{700}{\nano\metre} silicon nitride layer ($n=2.0$) lies on silica ($n=1.444$) with constant indices, and the \SI{20}{\nano\metre} Yee mesh of the production run is kept. Every solver receives the same per-component permittivity array of the pillar layer, evaluated at its own Yee positions. The reference is TorchFDTD on a CPU with the production tile contract (1,200 steps, 12 absorbing cells along $z$). The maps $T$ are the means of the exit-plane field over the 14$\times$14 pupil cells of \SI{0.25}{\micro\metre} inside the patch, each divided by the field of a homogeneous silica run of the same solver.

The solvers differ in three respects. The source is a soft plane source of the TorchFDTD band for the reference and Meep and a Gaussian-modulated pulse centered at \SI{580}{THz} with a \SI{100}{THz} width for FDTDX. The absorbing layers are 12 cells along $z$ in TorchFDTD and FDTDX and a \SI{0.24}{\micro\metre} PML along $z$ in Meep, all with periodic lateral faces. The normalization is the incident flux and field of the production reference for TorchFDTD, the flux per area of the homogeneous silica run for Meep, and the field of the homogeneous run of the same solver for FDTDX. Meep also uses the time convention $\exp(-i\omega t)$, a Courant number of 0.5 and no subpixel averaging of the given permittivity, and the phase signs of the three codes were matched before the comparison.

Table~\ref{tab:xsmall} lists the agreement with the reference. The metrics are the complex correlation, the relative $L^2$ difference, and the $|T|$-weighted RMS phase difference after removal of the global phase. The ranges cover the co-polarized maps $T_{xx}$ and $T_{yy}$ of the hologram and $T_{xx}$ of the lens, and the cross-polarized maps $T_{yx}$ and $T_{xy}$ of the hologram and $T_{yx}$ of the lens.

\begin{table}[tbp]
\centering
\small
\caption{Jones maps of the \SI{4}{\micro\metre} periodic patches against TorchFDTD on a CPU. Ranges are the minimum and maximum over the three co-polarized and the three cross-polarized maps of the hologram and the lens patches.}
\label{tab:xsmall}
\begin{tabular}{llccc}
\toprule
Solver & Maps & Correlation & Relative $L^2$ (\%) & Phase RMS (rad) \\
\midrule
Meep & co-polarized & 0.9996 to 0.9998 & 2.4 to 3.5 & 0.014 to 0.017 \\
Meep & cross-polarized & 0.9945 to 0.9957 & 9.6 to 10.7 & 0.064 to 0.078 \\
FDTDX & co-polarized & 1.0000 & 0.41 to 0.46 & 0.003 \\
FDTDX & cross-polarized & 0.9982 to 0.9989 & 4.7 to 5.9 & 0.032 to 0.042 \\
\bottomrule
\end{tabular}
\end{table}

\subsection*{Adjoint gradient of one tile against FDTDX}
\begin{figure}[tbp]
\centering
\includegraphics[width=\textwidth]{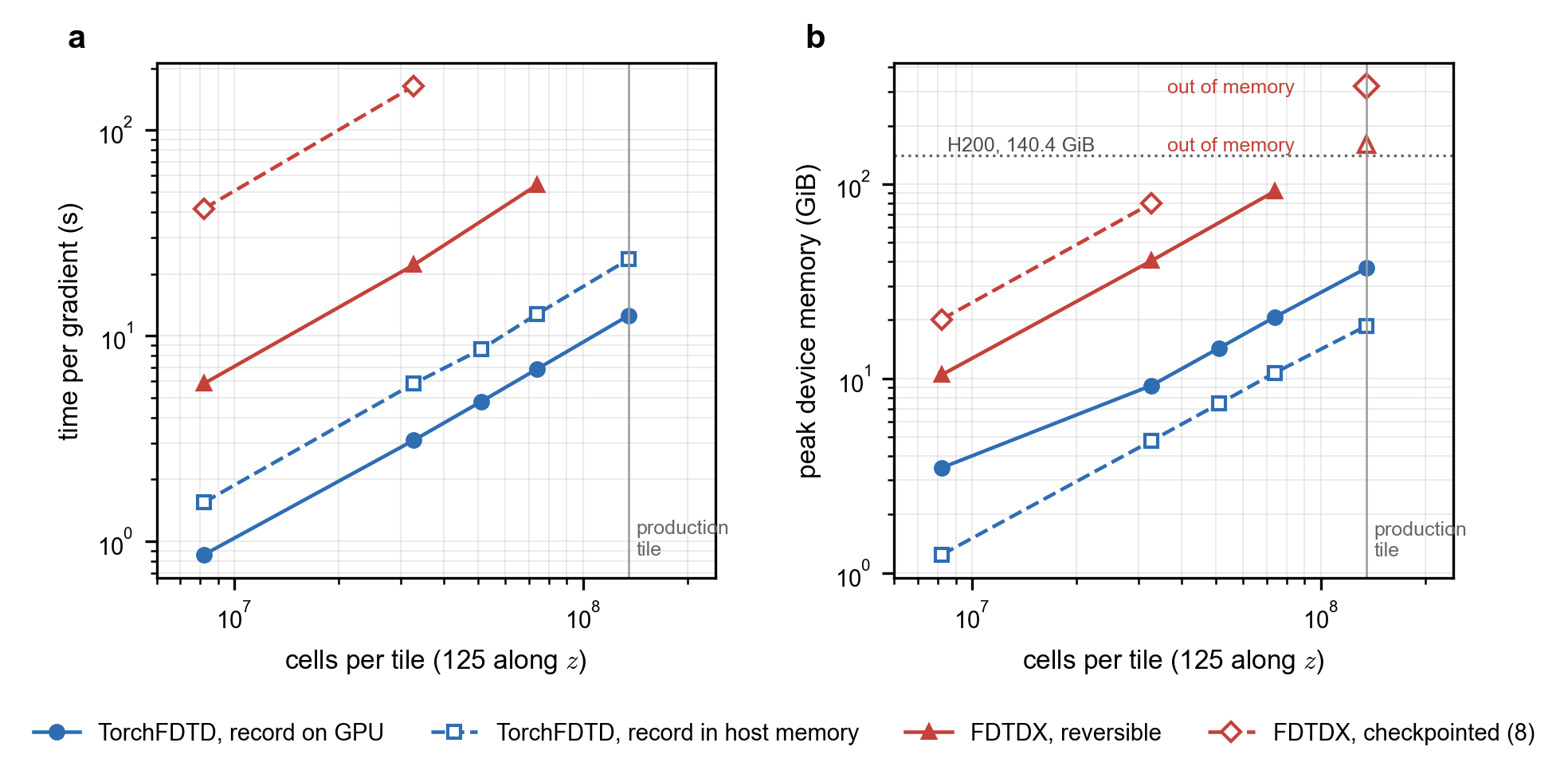}
\caption{\textbf{Adjoint gradient of one tile on one H200 GPU.} \textbf{a}, Time per gradient (forward and reverse pass, 1,200 steps, three wavelengths) against the number of cells of the window, for TorchFDTD with the recorded planes on the GPU or in host memory and for FDTDX in its reversible mode and with eight checkpoints. \textbf{b}, Peak device memory. Open markers are the allocations that FDTDX requested at the production window of $1040\times1040\times125$ cells (vertical line), which exceed the 141~GB of the device (dotted line).}
\label{fig:fdtdxadj}
\end{figure}

The gradient of one tile was computed by TorchFDTD and by the JAX solver FDTDX 0.6.2\cite{cpc_mahlau2024,cpc_mahlau2026joss} on one H200 GPU with 141~GB of memory (Supplementary Fig.~\ref{fig:fdtdxadj}). Both codes solve the same window of the polarization-switched hologram with $N\times N\times125$ cells of \SI{20}{\nano\metre}, the same diagonal anisotropic permittivity of the pillar layer, periodic lateral faces, 12 absorbing cells along $z$, an $x$-polarized plane source and 1,200 steps. The objective is the mean exit-plane intensity at 450, 540 and \SI{635}{\nano\metre}, normalized by the homogeneous run, and the gradient is taken with respect to the three permittivity components of the pillar layer. Times are medians of the forward and reverse passes without compilation, and memory is the peak of the device allocator. TorchFDTD kept the recorded planes either on the GPU or in pinned host memory, and FDTDX used its reversible mode and its checkpointed mode with eight checkpoints. At $N=256$, 512 and 768 the two gradients agree with cosine similarities of 0.99970, 0.99952 and 0.99925 and relative $L^{2}$ differences of 2.4, 3.1 and 3.9\%, and the two FDTDX modes agree with each other to $2\times10^{-7}$. At $N=768$, the largest window that fitted in the reversible mode of FDTDX, FDTDX took 53.9~s and 91.7~GiB and TorchFDTD 6.9~s and 20.7~GiB. At the production window, $N=1040$, TorchFDTD took 12.5~s and 37.0~GiB, or 23.5~s and 18.7~GiB with the planes in host memory, and both modes of FDTDX exceeded the device, requiring 160.5~GiB and 319~GiB. With nine wavelengths the TorchFDTD gradient of the production window took 12.6~s and 37.0~GiB. On the $64^{3}$ periodic dielectric box of a matched-discretization test (128 steps), the full permittivity gradient of TorchFDTD agrees with that of FDTDX 0.6.2 to $7.1\times10^{-8}$ in relative $L_2$ norm with checkpointing and to $1.0\times10^{-7}$ in its reversible mode on one H200 SXM GPU, and the two FDTDX modes agree with each other to $1.1\times10^{-7}$.

\subsection*{Full-aperture comparisons}
Each final layout is solved over its whole aperture with Tidy3D 2.10.2 on the same per-component permittivity array, and the near-field maps $T$ are compared with the production maps of TorchFDTD by the complex correlation and the relative $L^2$ difference after removal of one global complex factor. The objective of each design is evaluated on the maps of each solver with the same propagation. The objectives are listed in Supplementary Table~\ref{tab:freeformval} and the focal intensity of the lens and the hologram images are shown in Fig.~5f,g of the main text; this note gives the map and image metrics.

\begin{table}[tbp]
\centering
\small
\caption{\textbf{Designs scored by an independent solver.} Objective of each final binary layout, and of the meta-atom lens (last row), from TorchFDTD tiles with \SI{3.6}{\micro\metre} overlaps and from Tidy3D over the whole aperture, with the same evaluator. The correlation is that of the co-polarized exit-plane maps of Tidy3D and of the production tiles, and the image difference is the relative $L_{2}$ difference of the focal-plane (lens) or image-plane (holograms) intensities over the wavelengths or input polarizations.}
\label{tab:freeformval}
\begin{tabular}{lccccc}
\toprule
Design & TorchFDTD $J$ & Tidy3D $J$ & Ratio & Map corr. & Image diff. \\
\midrule
E1, lens & 0.1196 & 0.1179 & 0.986 & 0.997 & 1.5--2.7\% \\
E2, color hologram & 0.6730 & 0.6626 & 0.985 & 0.998 & 1.7--2.5\% \\
E3, pol.-switched hologram & 0.8439 & 0.8309 & 0.985 & 0.998 & 2.1\% \\
\addlinespace
E1, meta-atom lens & 0.0728 & 0.0722 & 0.992 & -- & -- \\
\bottomrule
\end{tabular}
\end{table}

\begin{figure}[tbp]
\centering
\includegraphics[width=\textwidth]{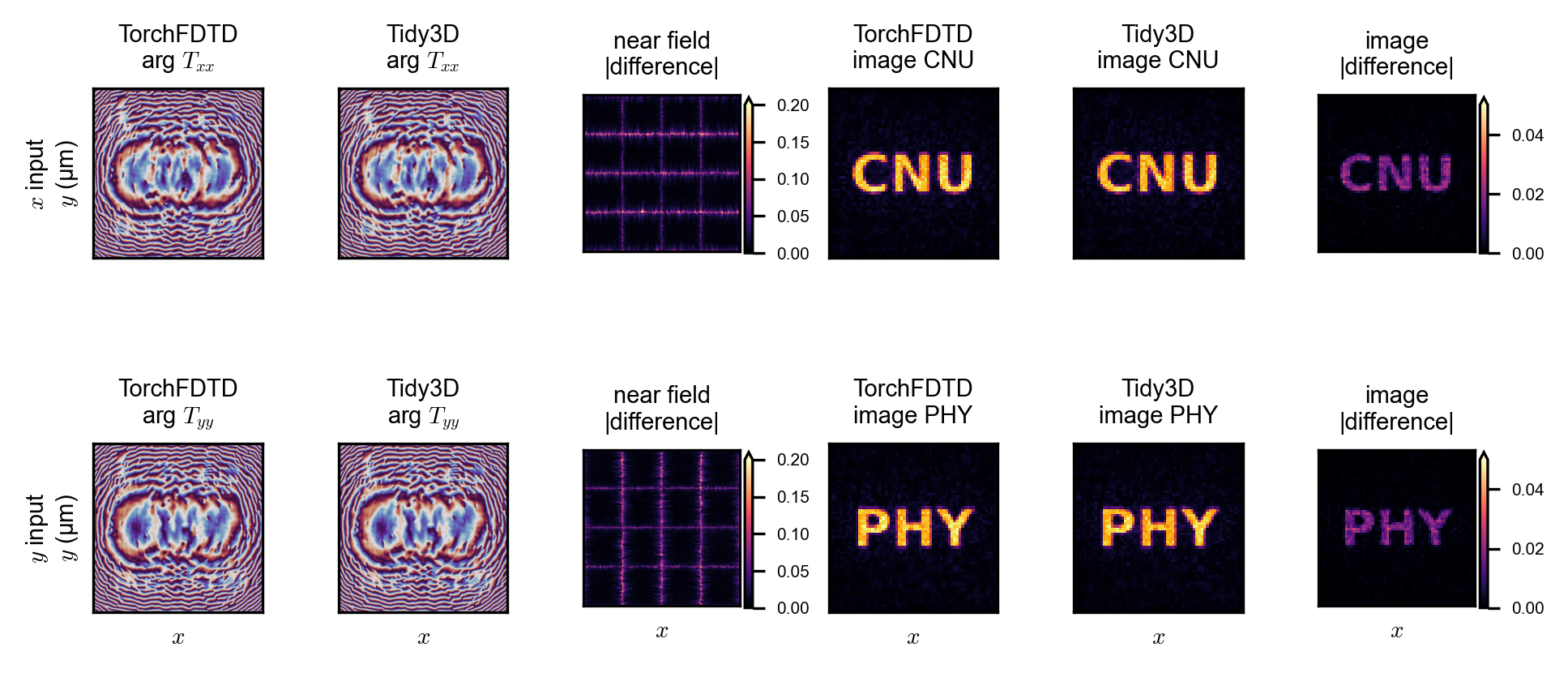}
\caption{\textbf{Polarization-switched hologram solved by TorchFDTD and by Tidy3D.} Rows: $x$ input with the map $T_{xx}$ and the image CNU, and $y$ input with $T_{yy}$ and the image PHY, at \SI{540}{\nano\metre}. From left: phase of the exit-plane map from the production TorchFDTD tiles and from Tidy3D over the whole \SI{100}{\micro\metre} aperture (after one global complex factor), magnitude of their difference relative to the maximum of $|T|$, image-plane intensity at $z=\SI{200}{\micro\metre}$ over the \SI{80}{\micro\metre} window from each solver, and the magnitude of their difference relative to the TorchFDTD peak.}
\label{fig:e3tidy}
\end{figure}

\emph{Polarization-switched hologram.} The binary layout of step 189 was solved over the \SI{100}{\micro\metre} aperture for $x$ and $y$ input at \SI{540}{\nano\metre}. The co-polarized maps $T_{xx}$ and $T_{yy}$ agree with correlations of 0.998 and 0.999 and relative $L^2$ differences of 5.9\% and 5.5\%, and the cross-polarized maps with correlations of 0.988 and relative differences of 15\%. The difference of the near field lies on the seams of the \SI{0.9}{\micro\metre}-overlap tiles (Supplementary Fig.~\ref{fig:e3tidy}). In the image plane at $z=\SI{200}{\micro\metre}$ the intensities differ by 2.1\% and 2.1\% in relative $L^2$ norm, with correlations of 0.99998. The shape accuracies $\hat S$ are 0.9751 and 0.9753 from TorchFDTD and 0.9750 and 0.9752 from Tidy3D, and the similarities to the image of the other polarization are 0.527 and 0.501 from TorchFDTD and 0.528 and 0.501 from Tidy3D. The objective is 0.8479 from the TorchFDTD run maps, 0.8439 from the tiles with \SI{3.6}{\micro\metre} overlaps and 0.8309 from Tidy3D, ratios of 0.980 and 0.985, and the window efficiency is 0.853 from Tidy3D against 0.871.

\emph{Broadband lens.} The binary layout of step 248 was solved over the \SI{200}{\micro\metre} aperture with the D4 quarter at the nine design wavelengths. The co-polarized maps agree with a correlation of 0.997 and a relative $L^2$ difference of 7.5\%, the cross-polarized maps with a correlation of 0.955, and the focal-plane intensities differ by 1.5 to 2.7\%. The objective is 0.1205 from the TorchFDTD run maps, 0.1196 from the tiles with \SI{3.6}{\micro\metre} overlaps and 0.1179 from Tidy3D, ratios of 0.979 and 0.986.

\emph{Color hologram.} The binary layout of step 158 was solved over the \SI{150}{\micro\metre} aperture for $x$ input at 450, 540 and \SI{635}{\nano\metre}. The co-polarized maps agree with correlations of 0.998, 0.998 and 0.998 and relative $L^2$ differences of 6.3\%, 6.5\% and 7.1\%. In the image plane at $z=\SI{300}{\micro\metre}$ the intensities of the three channels differ by 2.5\%, 1.9\% and 1.7\% in relative $L^2$ norm. The objective is 0.6761 from the TorchFDTD run maps, 0.6730 from the tiles with \SI{3.6}{\micro\metre} overlaps and 0.6626 from Tidy3D, ratios of 0.980 and 0.985, and the efficiencies into the window are 0.724, 0.785 and 0.758 from Tidy3D against 0.742, 0.799 and 0.769.

\subsection*{Single-domain solutions against production tiling}
\begin{figure}[tbp]
\centering
\includegraphics[width=\textwidth]{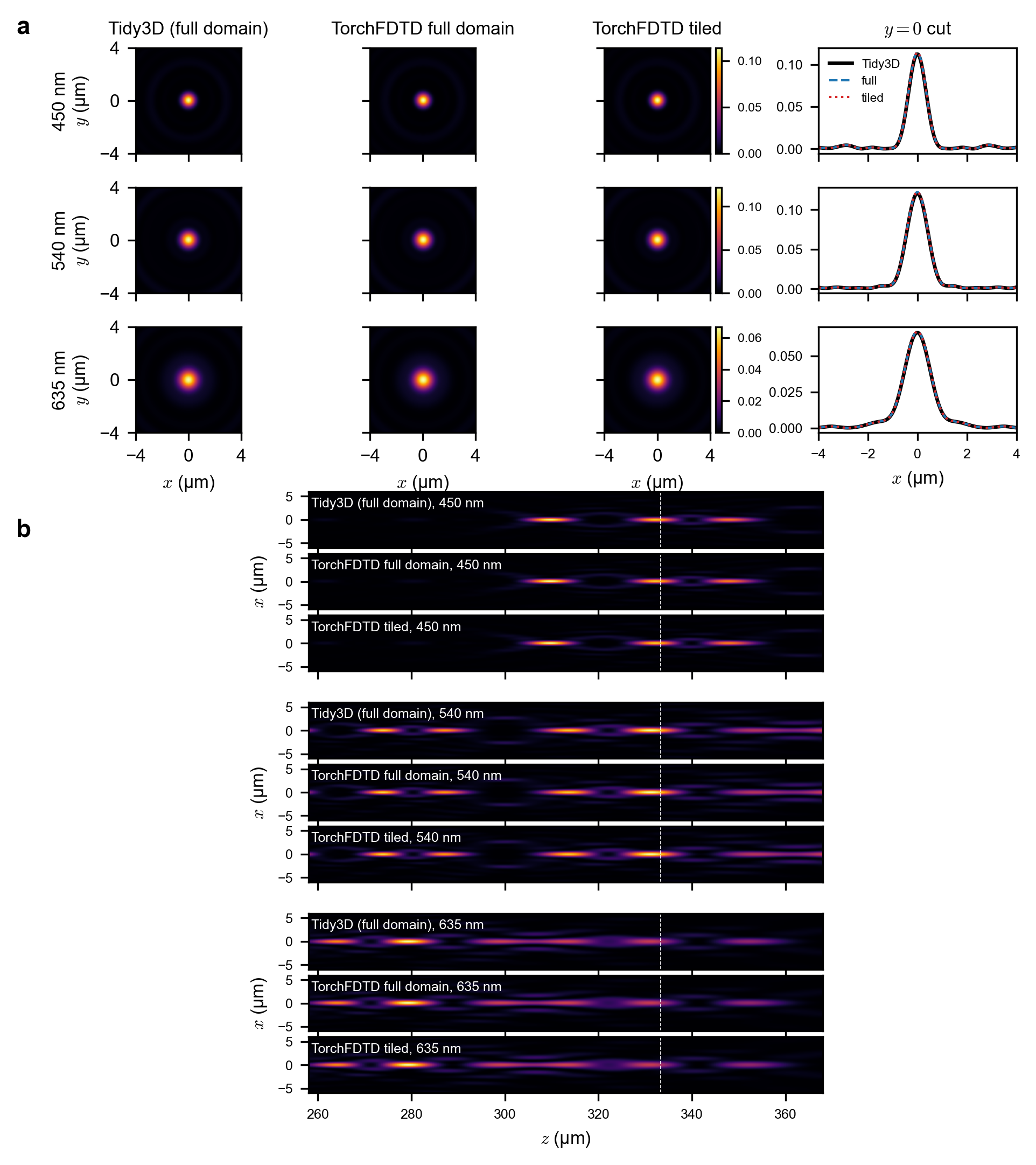}
\caption{\textbf{Single-domain and tiled solutions of the \SI{200}{\micro\metre} lens.} \textbf{a}, Focal-plane intensity normalized by the peak of an ideal lossless lens, from Tidy3D over the whole aperture, from one TorchFDTD domain over the whole aperture and from the production tiles with \SI{3.6}{\micro\metre} overlaps, with the cut at $y=0$. Each row shares one color scale. \textbf{b}, Intensity in the plane $y=0$ from $z=258$ to \SI{368}{\micro\metre} with equal axis scales. The dashed line marks the design focus $f=\SI{333.3}{\micro\metre}$.}
\label{fig:threeway}
\end{figure}

The \SI{200}{\micro\metre} meta-atom lens of the main text (post library, $x$ input, constant indices, 450, 540 and \SI{635}{\nano\metre}) was solved three ways on the same per-component permittivity array: as one TorchFDTD domain divided into slabs with the D4 quarter (1,200 steps), as one Tidy3D 2.10.2 domain with the same quarter symmetry, and by the production tiles with \SI{0.9}{\micro\metre} and \SI{3.6}{\micro\metre} overlaps. Each near-field map is normalized by the homogeneous run of its own solver. The focal fields follow from the same propagation and the same ideal-lens normalization for all maps. Supplementary Fig.~\ref{fig:threeway} shows the focal intensity and the $x$--$z$ intensity in the plane $y=0$, and Supplementary Table~\ref{tab:threeway} lists the relative RMS differences averaged over the three wavelengths. The near-field difference is taken after removal of one global complex factor per wavelength. With \SI{0.9}{\micro\metre} overlaps it falls from 13.4\% within \SI{0.5}{\micro\metre} of a seam to 0.9\% near the core centers, and with \SI{3.6}{\micro\metre} overlaps from 2.9\% to below 0.1\% (Fig.~2e of the main text). The focal difference is taken on the $12\times\SI{12}{\micro\metre}$ focal-plane intensity without rescaling. The on-axis peaks lie at 310.8, 330.8 and \SI{278.3}{\micro\metre} for all four solutions, and the focusing efficiencies at 450, 540 and \SI{635}{\nano\metre} are 0.1042, 0.1303 and 0.0906 for the single TorchFDTD domain, 0.1042, 0.1303 and 0.0906 for the \SI{3.6}{\micro\metre} tiles and 0.1025, 0.1286 and 0.0896 for Tidy3D.

\begin{table}[tbp]
\centering
\small
\caption{Relative RMS differences between the solutions of the \SI{200}{\micro\metre} lens, mean over 450, 540 and \SI{635}{\nano\metre}. Near field: complex map $T_{xx}$ on the \SI{0.25}{\micro\metre} pupil cells. Focus: intensity on the focal plane.}
\label{tab:threeway}
\begin{tabular}{lcc}
\toprule
Pair & Near field (\%) & Focus (\%) \\
\midrule
TorchFDTD single domain vs Tidy3D & 2.20 & 1.44 \\
TorchFDTD tiles, \SI{3.6}{\micro\metre} overlap, vs Tidy3D & 3.00 & 1.44 \\
TorchFDTD tiles, \SI{0.9}{\micro\metre} overlap, vs Tidy3D & 6.35 & 1.27 \\
TorchFDTD tiles, \SI{3.6}{\micro\metre} overlap, vs single domain & 1.86 & 0.02 \\
TorchFDTD tiles, \SI{0.9}{\micro\metre} overlap, vs single domain & 5.83 & 0.26 \\
\bottomrule
\end{tabular}
\end{table}

The three optimized designs were also solved as single TorchFDTD domains over their whole apertures, the lens as its D4 quarter, with the slab decomposition, 1,200 steps and the permittivity arrays of the production tiles, and scored by the same evaluator (Supplementary Table~\ref{tab:singledomain}). The single-domain objectives agree with those of the tiles with \SI{3.6}{\micro\metre} overlaps to within 0.01\%. The co-polarized near fields differ by 1.6 to 2.5\% and the focal-plane intensities and hologram images by at most $8\times10^{-4}$ in relative $L_{2}$ norm. For the color hologram, tiles with the \SI{0.9}{\micro\metre} overlaps of the optimization give $J_{2}=0.6761$, near-field differences of 5.8 to 6.9\% and image differences of 0.6 to 0.9\%.

\begin{table}[tbp]
\centering
\small
\caption{Optimized designs solved as single domains. Objective $J$ of each final binary layout from one TorchFDTD domain, from TorchFDTD tiles with \SI{3.6}{\micro\metre} overlaps and from Tidy3D, with the same evaluator. Near field: relative $L_{2}$ difference of the co-polarized exit-plane maps between the single domain and the tiles over the design wavelengths or input polarizations. Image: relative $L_{2}$ difference of the focal-plane intensity (lens) or image-plane intensity (holograms) between the single domain and the tiles.}
\label{tab:singledomain}
\footnotesize
\setlength{\tabcolsep}{2.5pt}
\begin{tabular}{lcccccc}
\toprule
Design & Grid (cells) & Single $J$ & Tiled $J$ & Tidy3D $J$ & Near field (\%) & Image \\
\midrule
E1 (quarter) & $5113^{2}\times125$ & 0.11962 & 0.11961 & 0.11791 & 2.0--2.5 & $(0.9\text{--}2.4)\times10^{-4}$ \\
E2 & $7724^{2}\times125$ & 0.67300 & 0.67297 & 0.66264 & 1.8--2.1 & $(4.5\text{--}8.4)\times10^{-4}$ \\
E3 & $5224^{2}\times125$ & 0.84395 & 0.84392 & 0.83085 & 1.6 & $4.6\times10^{-4}$ \\
\bottomrule
\end{tabular}
\end{table}

\section{Sensitivity to layer thickness and edge bias}
\label{si:robust}
\begin{figure}[tbp]
\centering
\includegraphics[width=\textwidth]{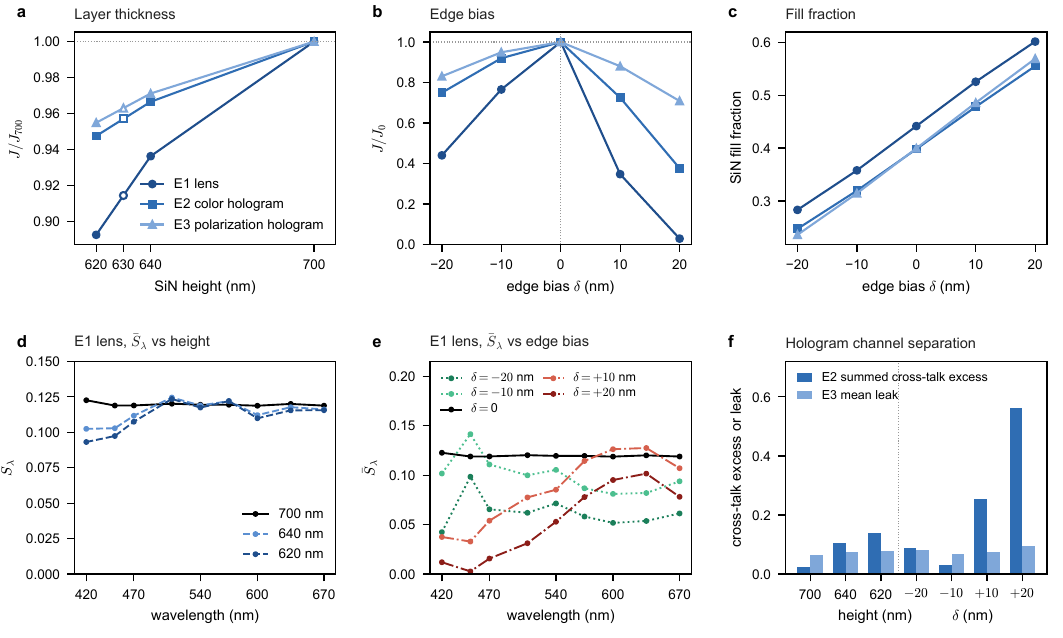}
\caption{\textbf{Layer thickness and edge bias.} Final binary freeform layouts scored by the TorchFDTD tiles with \SI{3.6}{\micro\metre} overlaps. (a) Objective relative to the \SI{700}{\nano\metre} layer for layers of 620 and \SI{640}{\nano\metre}; open markers, their mean as the \SI{630}{\nano\metre} estimate. (b) Objective relative to the unbiased layout against the edge bias $\delta$ at \SI{700}{\nano\metre} height; $\delta>0$ widens the silicon nitride features (overdose of a negative-tone exposure). (c) Silicon nitride fill fraction inside the aperture. (d, e) Focal intensity of the lens relative to the ideal lens, mean of $x$ and $y$ input, for the three heights (d) and the five biases (e). (f) Summed cross-talk excess of the color hologram and leak of the polarization-switched hologram, mean of $x$ and $y$ input.}
\label{fig:robust}
\end{figure}

The three final binary freeform layouts were scored again with two fabrication changes: a thinner silicon nitride layer, and a uniform lateral shift of every layout edge such as an over- or underdosed electron-beam exposure produces. The evaluator is the one of the main-text scores, TorchFDTD tiles with \SI{15}{\micro\metre} cores and \SI{3.6}{\micro\metre} overlaps on the \SI{20}{\nano\metre} mesh with the material averaging and reference solutions of Methods. Only the permittivity of the layer changes. Evaluated again at \SI{700}{\nano\metre}, the unchanged layouts reproduce the reported objectives to $4\times10^{-10}$. Supplementary Fig.~\ref{fig:robust} shows the results and Supplementary Table~\ref{tab:robust} lists them.

\emph{Layer thickness.} The layer height enters the \SI{20}{\nano\metre} Yee lattice as whole cells along $z$, 35 cells for \SI{700}{\nano\metre}, with no sub-cell averaging along $z$. A \SI{630}{\nano\metre} layer is therefore not representable on this lattice. We evaluate the neighboring heights of 620 and \SI{640}{\nano\metre} (31 and 32 cells) and quote their mean as the \SI{630}{\nano\metre} estimate. At \SI{630}{\nano\metre} the objectives are 0.914, 0.957 and 0.963 of their \SI{700}{\nano\metre} values for the lens, the color hologram and the polarization-switched hologram. For the lens the change sits at the blue end of the band: at \SI{620}{\nano\metre}, $\bar S_{\lambda}$ at 420 and \SI{450}{\nano\metre} falls from 0.123 and 0.119 to 0.093 and 0.097, while from 470 to \SI{670}{\nano\metre} it stays between 0.107 and 0.124. For the color hologram the summed cross-talk excess rises from 0.022 to 0.104 at \SI{640}{\nano\metre} and 0.137 at \SI{620}{\nano\metre}, with similarities of 0.887, 0.856 and 0.886 at \SI{620}{\nano\metre} against 0.911, 0.856 and 0.879. For the polarization-switched hologram the leaks for $x$ and $y$ input rise from 0.059 and 0.073 to 0.071 and 0.084 at \SI{620}{\nano\metre}, and the similarities stay at 0.972 and 0.975.

\emph{Edge bias.} We model a dose error as a uniform edge bias $\delta$ of the silicon nitride pattern at the fixed height of \SI{700}{\nano\metre}. The convention assumes a negative-tone resist such as hydrogen silsesquioxane, exposed where silicon nitride remains, so an overdose widens the silicon nitride features ($\delta>0$) and an underdose narrows them ($\delta<0$). The bias is applied to each layout on its \SI{10}{\nano\metre} design grid by binary dilation ($\delta>0$) or erosion ($\delta<0$) with a disk of radius $|\delta|$, for $\delta=\pm10$ and $\pm\SI{20}{\nano\metre}$. The evaluator gives each electric-field component of the \SI{20}{\nano\metre} mesh the mean of the $2\times2$ design pixels centered on it, so a \SI{10}{\nano\metre} shift of a straight edge changes the fill of the adjacent cells by 0.5 and enters through the sub-cell material averaging. In the unbiased layouts 11 to 13\% of the \SI{20}{\nano\metre} cells inside the aperture are partly filled. A bias of \SI{10}{\nano\metre} changes 7.9 to 8.6\% of the design pixels and the fill fraction by about 0.08: from 0.442 to 0.358 and 0.525 for the lens, from 0.398 to 0.320 and 0.478 for the color hologram and from 0.400 to 0.315 and 0.486 for the polarization-switched hologram, at $\delta=-10$ and $+\SI{10}{\nano\metre}$.

For all three designs a narrowing bias lowers the objective less than a widening bias of the same size, and the holograms respond less than the lens. At $\delta=-10$ and $+\SI{10}{\nano\metre}$ the objectives are 0.766 and 0.347 of their unbiased values for the lens, 0.920 and 0.725 for the color hologram and 0.949 and 0.880 for the polarization-switched hologram, and at $\delta=-20$ and $+\SI{20}{\nano\metre}$ they are 0.440 and 0.029, 0.750 and 0.376, and 0.831 and 0.709. The lens responds with a shift of its band. At $\delta=+\SI{10}{\nano\metre}$, $\bar S_{\lambda}$ at 420 and \SI{450}{\nano\metre} falls from 0.123 and 0.119 to 0.037 and 0.033, while at 600 and \SI{635}{\nano\metre} it is 0.126 and 0.127. At $\delta=-\SI{10}{\nano\metre}$ it rises to 0.141 at \SI{450}{\nano\metre} and falls to 0.081 to 0.094 from 570 to \SI{670}{\nano\metre}. Because $J_{1}$ is a power mean with $q=-8$, the weakest wavelength sets these values. In the color hologram the bias opens cross-talk: the summed cross-talk excess is 0.031 and 0.255 at $\delta=-10$ and $+\SI{10}{\nano\metre}$ and 0.088 and 0.563 at $\mp\SI{20}{\nano\metre}$, and at $\delta=+\SI{20}{\nano\metre}$ the similarities are 0.656, 0.774 and 0.836. In the polarization-switched hologram the leaks for $x$ and $y$ input are 0.062 and 0.075 at $\delta=-\SI{10}{\nano\metre}$, 0.067 and 0.083 at $+\SI{10}{\nano\metre}$, 0.075 and 0.088 at $-\SI{20}{\nano\metre}$ and 0.085 and 0.106 at $+\SI{20}{\nano\metre}$, and the similarities stay between 0.923 and 0.967.

\begin{table}[tbp]
\centering
\small
\caption{Objective of the final binary freeform layouts relative to the \SI{700}{\nano\metre} unbiased layout ($J/J_{0}$) for layer heights $H$ and edge biases $\delta$, and the silicon nitride fill fraction inside the aperture. The \SI{630}{\nano\metre} column is the mean of the 620 and \SI{640}{\nano\metre} results.}
\label{tab:robust}
\footnotesize
\setlength{\tabcolsep}{3pt}
\begin{tabular}{lcccccccc}
\toprule
 & $J_{0}$ & \multicolumn{3}{c}{$H$ (nm), $\delta=0$} & \multicolumn{4}{c}{$\delta$ (nm), $H=\SI{700}{\nano\metre}$} \\
Design & & 620 & 640 & 630 & $-20$ & $-10$ & $+10$ & $+20$ \\
\midrule
E1, $J/J_{0}$ & 0.1196 & 0.893 & 0.936 & 0.914 & 0.440 & 0.766 & 0.347 & 0.029 \\
E2, $J/J_{0}$ & 0.6730 & 0.948 & 0.967 & 0.957 & 0.750 & 0.920 & 0.725 & 0.376 \\
E3, $J/J_{0}$ & 0.8439 & 0.955 & 0.971 & 0.963 & 0.831 & 0.949 & 0.880 & 0.709 \\
\midrule
E1, fill (0.442) & & & & & 0.283 & 0.358 & 0.525 & 0.601 \\
E2, fill (0.398) & & & & & 0.248 & 0.320 & 0.478 & 0.555 \\
E3, fill (0.400) & & & & & 0.236 & 0.315 & 0.486 & 0.569 \\
\bottomrule
\end{tabular}
\end{table}

\section{Cost accounting}
\label{si:cost}
Each production run records the wall time of every stage of every iteration: setup, reference solutions, tile forward passes, recorded forward passes, tile adjoints, assembly and propagation, optimizer, binary evaluation, file output, collective communication and idle time. Peak Torch allocation and reservation, host resident memory and sampled device utilization are stored separately. The cost of a design is the wall-clock time of its optimization steps multiplied by the hourly price of the node. Setup, including the reference flux, takes 66~s for the lens and about 6~s when a run continues from a checkpoint. The node leases totalled 5.04~h, or \$185, for the lens, 10.1~h, or \$282, for the color hologram and 12.4 to 12.6~h, or \$340 to \$347, for the polarization-switched hologram. The gradient of one full aperture from 50~\textmu m to 1~mm was measured on four H100 GPUs, with every tile solved up to \SI{200}{\micro\metre} and three tiles per GPU solved and the rest extrapolated at \SI{400}{\micro\metre} and 1~mm (Fig.~6d of the main text). Supplementary Table~\ref{tab:scaling} summarizes how one gradient scales with the aperture diameter.

On eight H200 GPUs a step of the lens took a median of 75.6~s and a mean of 67.8~s, because the binary layout is evaluated at every second step (54.0~s without and 76.0~s with that evaluation). On average, tile forward solutions took 23.8~s of a step, tile adjoint solutions 30.5~s, binary evaluation 11.3~s, file output 1.3~s, assembly, propagation and objective 0.36~s, the optimizer 0.33~s and communication 0.08~s, and a GPU was without recorded work for 3.6\% of the step. The polarization-switched hologram took a median of 225.3~s and a mean of 201.6~s per step on six H200 GPUs, of which 54.4~s were tile forward and 119.2~s tile adjoint solutions. The color hologram took a median of 222.0~s per step for its first 10 steps and 217.5~s from step 10 to step 158, on the nodes given in Methods, with on average 57.9~s of tile forward and 119.3~s of tile adjoint solutions. On four H100 GPUs, one full-aperture gradient with three wavelengths took 88~s at \SI{50}{\micro\metre}, 198~s at \SI{100}{\micro\metre} and 627~s at \SI{200}{\micro\metre}, and the extrapolation gives 2{,}154~s at \SI{400}{\micro\metre} and 12{,}464~s for the 2{,}277 tiles of 1~mm. The first GPU, which also holds the global \SI{10}{\nano\metre} design grid and the pupil arrays, reaches 66~GiB at 1~mm, and the host memory of the node 307~GiB.

\begin{table}[tbp]
\centering
\small
\caption{\textbf{Scaling of one gradient with the aperture diameter $D$.} Fixed mesh, layer thickness and numerical aperture. The direct adjoint solves the focal volume by FDTD and stores the forward history; this work solves the layer by FDTD, reconstructs its interior backwards, records only the planes that bound the absorbers and propagates the exit field by fast Fourier transforms.}
\label{tab:scaling}
\begin{tabular}{lcc}
\toprule
 & Direct time-domain adjoint & This work \\
\midrule
FDTD cells & $D^{3}$ & $D^{2}$ \\
Time steps & $D$ & constant \\
Computation & $D^{4}$ & $D^{2}\log D$ \\
Gradient memory, total & $D^{4}$ & $D^{2}$ \\
Gradient memory per GPU & $D^{4}$ & constant (39.7~GiB per $1040^{2}\times125$ tile) \\
\bottomrule
\end{tabular}
\end{table}

\section{Prior large-area designs and the memory of one gradient}
\label{si:prior}
\begin{figure}[tbp]
\centering
\includegraphics[width=\linewidth]{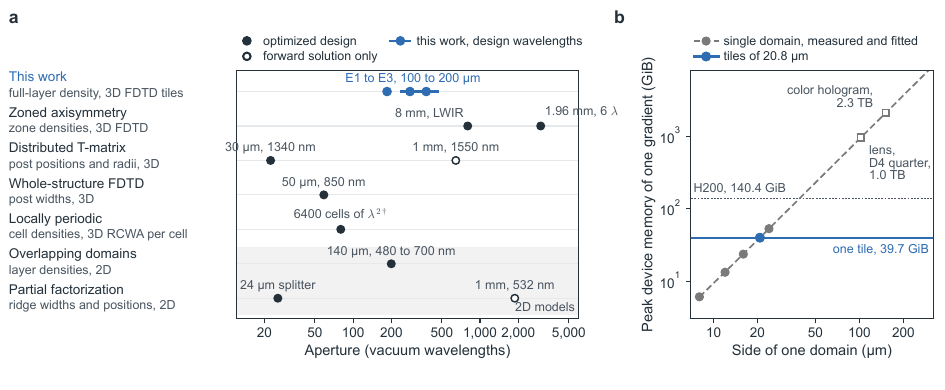}
\caption{Prior large-area designs and the memory of one gradient. (a) Apertures of full-wave inverse designs of large-area metasurfaces, in vacuum wavelengths, grouped by method and by the kind of design variable: zoned axisymmetry\cite{ncs_zda2025}, distributed T-matrix\cite{ncs_skarda2022}, whole-structure FDTD\cite{ncs_mansouree2021}, locally periodic\cite{ncs_lpa2019}, overlapping domains\cite{cpc_lin2019} and partial factorization\cite{ncs_apf2022,ncs_apfinverse2024}. Filled markers are optimized designs, open markers are forward solutions only, and the shaded band marks two-dimensional electromagnetic models. The dagger marks the square-equivalent width of the 6,400 cells of the locally periodic design. The blue markers are the three designs of this work, 100 to \SI{200}{\micro\metre} across, each spanning its design wavelengths. (b) Peak device memory of one forward and reverse call of a single domain against the side of the domain, measured on domains of 8 to \SI{24}{\micro\metre} at the \SI{20}{\nano\metre} mesh with 1,200 steps and three wavelengths (gray circles) and extrapolated by the fitted linear model (dashed line) to the D4 quarter of the \SI{200}{\micro\metre} lens, 1.0~TB, and to the full \SI{150}{\micro\metre} color hologram, 2.3~TB (open squares). The dotted line marks the memory of one H200. The \SI{20.8}{\micro\metre} point is one production tile, whose 39.7~GiB does not depend on the aperture (blue line).}
\label{fig:landscape}
\end{figure}

Supplementary Fig.~\ref{fig:landscape} places the three designs among earlier large-area full-wave inverse designs and compares the device memory of a single-domain gradient with that of one tile. The single-domain memory was measured on domains of 8 to \SI{24}{\micro\metre} at the production mesh and grows linearly with the number of cells and of time steps, so the fitted model extrapolates it to the production apertures.

\section{Final binary layouts}
\label{si:layouts}
\begin{figure}[p]
\centering
\includegraphics[width=\textwidth]{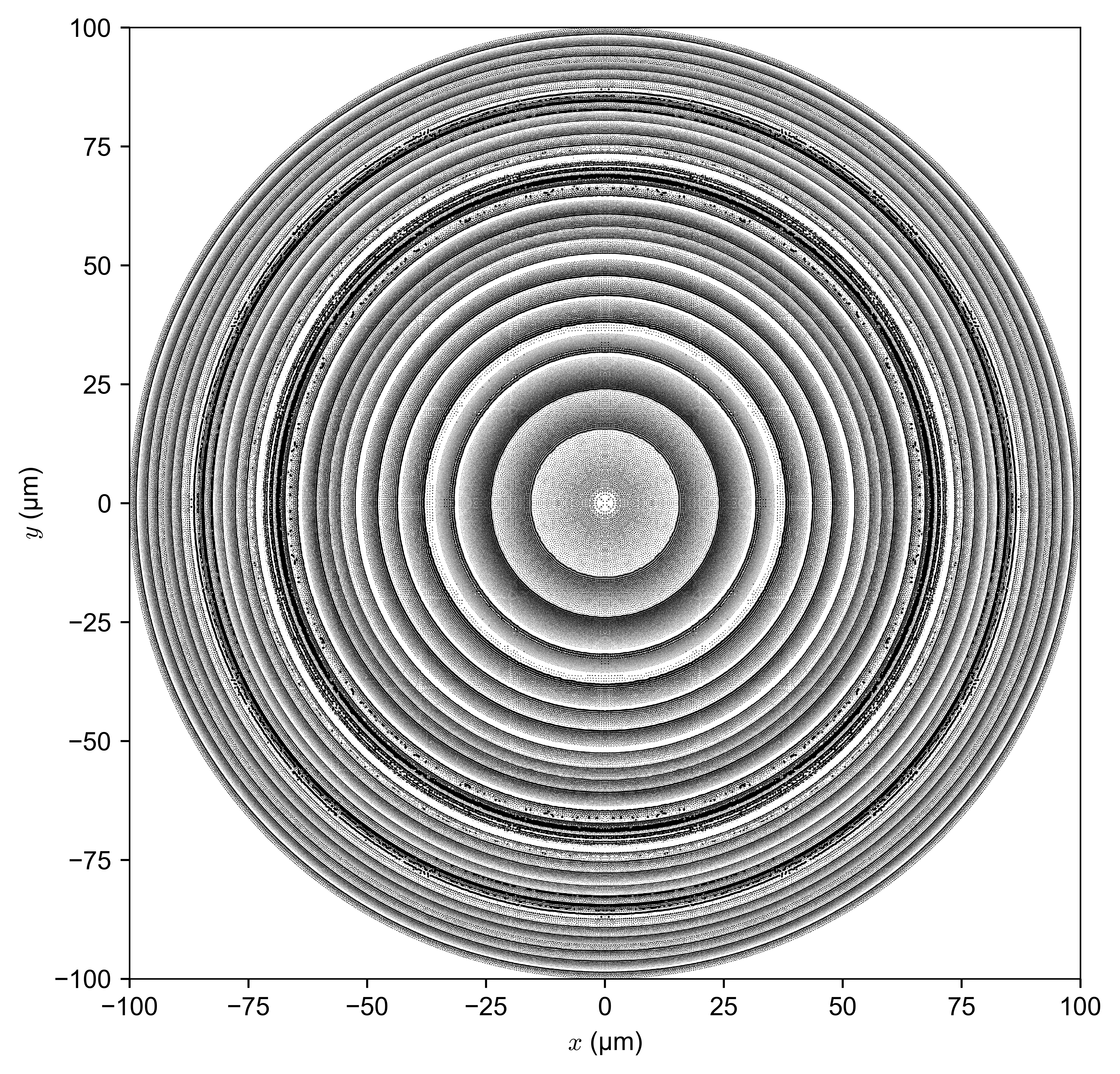}
\caption{\textbf{Final binary layout of the broadband lens.} Step 248, \SI{200}{\micro\metre} aperture, D4 symmetric, shown as the fill fraction on 40~nm blocks of the 10~nm raster.}
\label{fig:topview_e1}
\end{figure}

\begin{figure}[p]
\centering
\includegraphics[width=\textwidth]{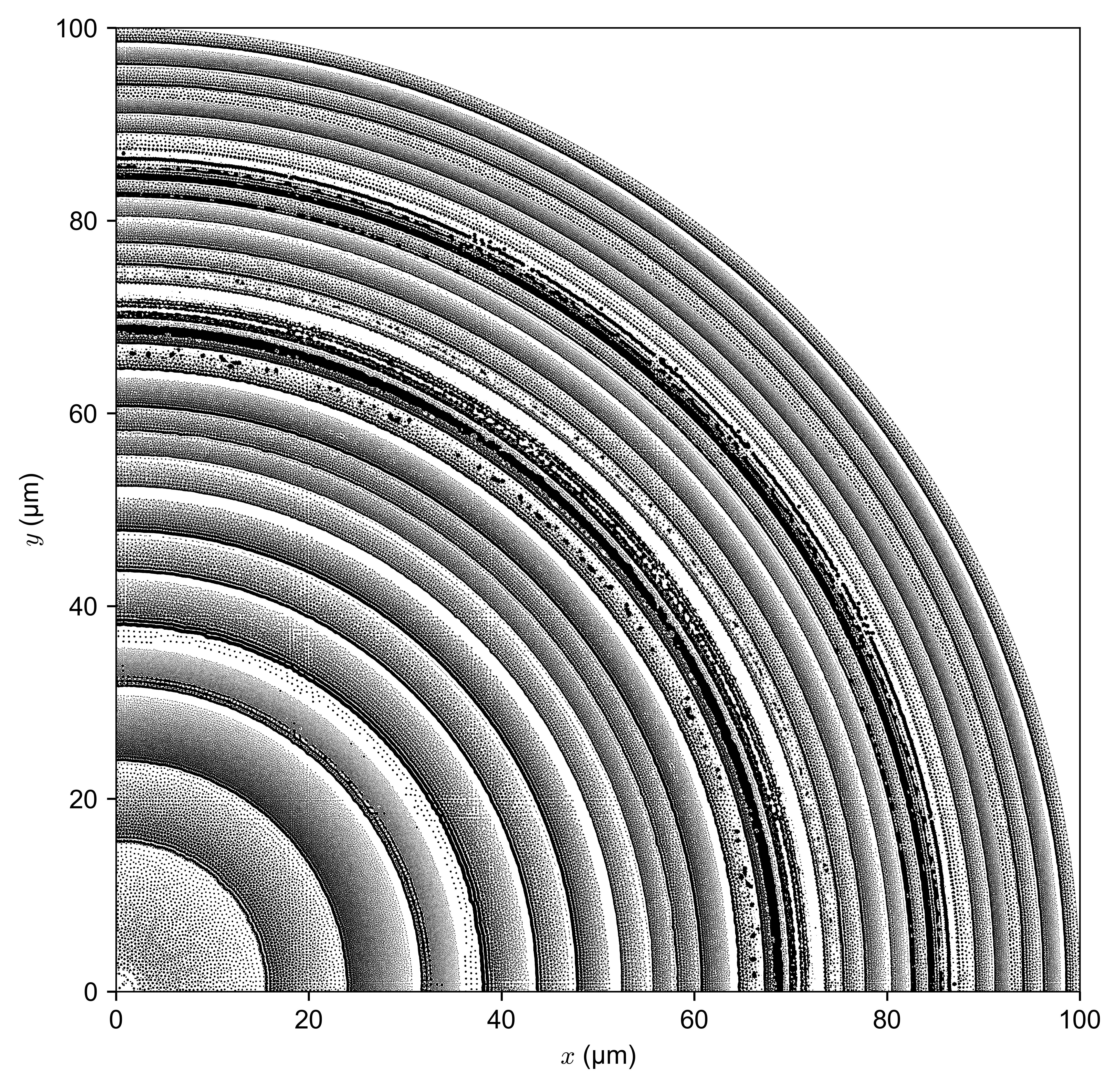}
\caption{\textbf{One quadrant of the lens layout.} The quadrant $x,y\geq0$ of Supplementary Fig.~\ref{fig:topview_e1}, fill fraction on 20~nm blocks.}
\label{fig:topview_e1quad}
\end{figure}

\begin{figure}[p]
\centering
\includegraphics[width=\textwidth]{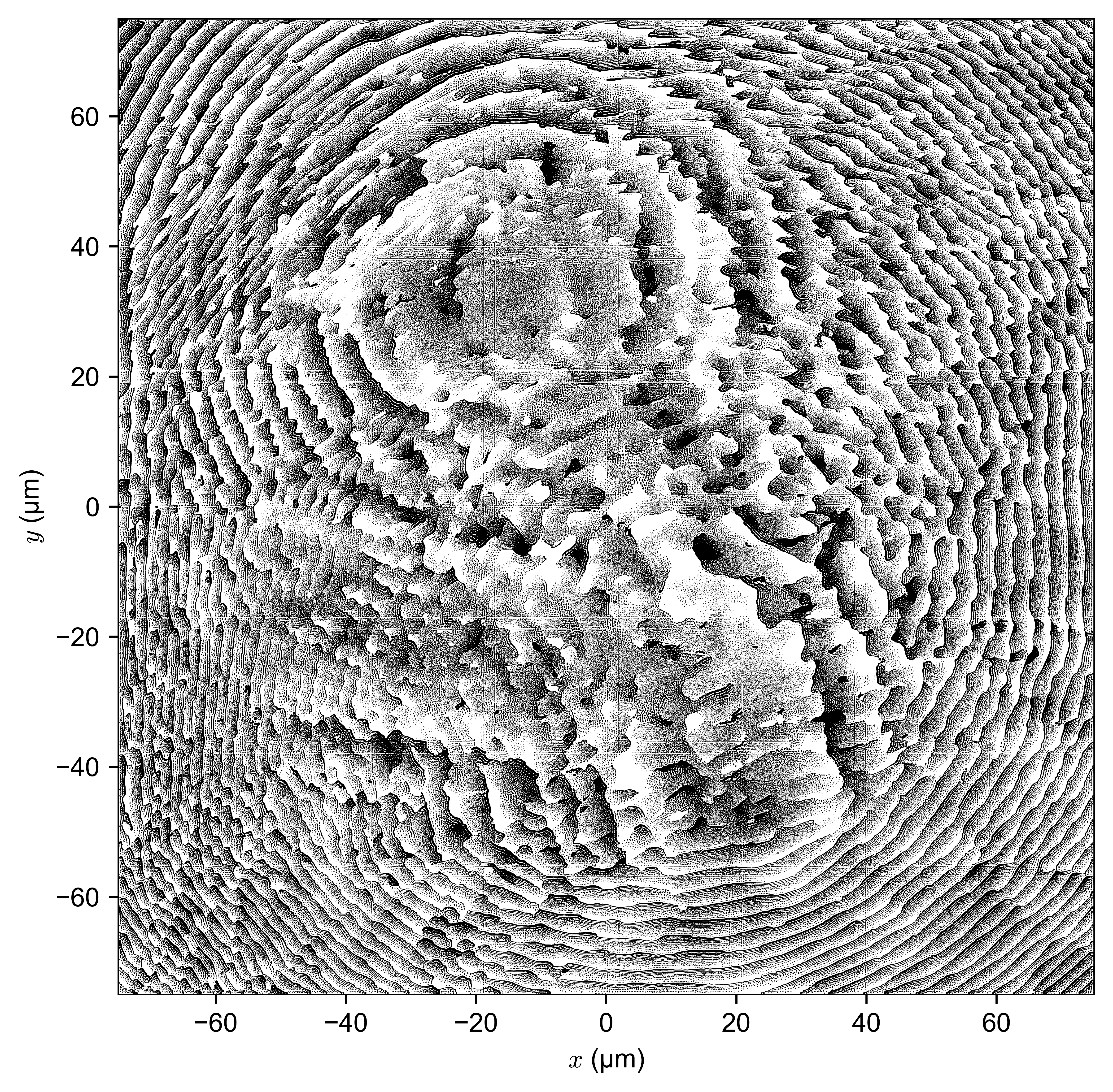}
\caption{\textbf{Final binary layout of the color hologram.} Step 158, \SI{150}{\micro\metre} square aperture, fill fraction on 30~nm blocks.}
\label{fig:topview_e2}
\end{figure}

\begin{figure}[p]
\centering
\includegraphics[width=\textwidth]{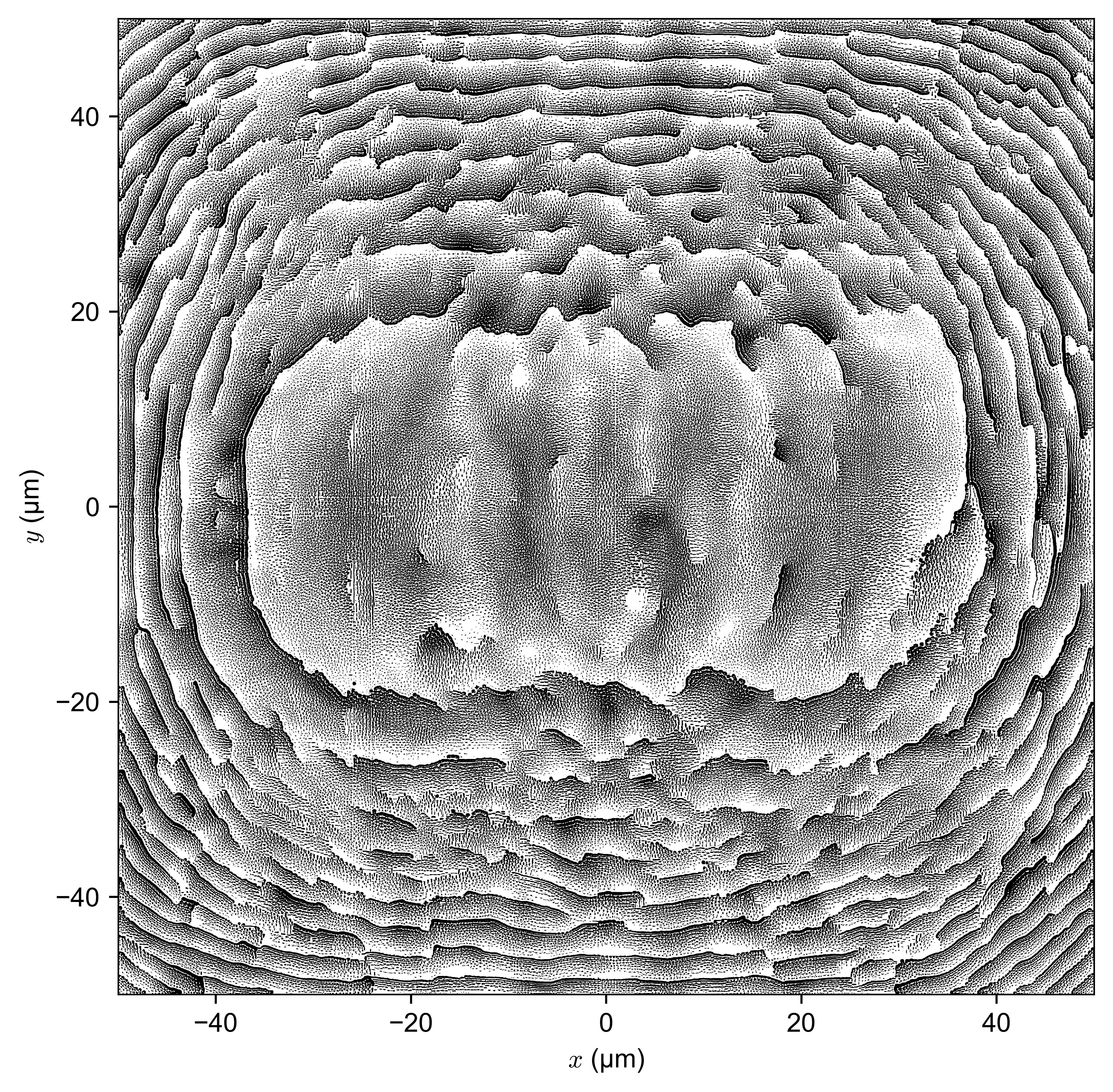}
\caption{\textbf{Final binary layout of the polarization-switched hologram.} Step 189, \SI{100}{\micro\metre} square aperture, fill fraction on 20~nm blocks.}
\label{fig:topview_e3}
\end{figure}

Supplementary Figs.~\ref{fig:topview_e1} to \ref{fig:topview_e3} show the final binary layouts of the three designs at the full page width. Silicon nitride is black. The 10~nm design rasters are shown as fill fractions averaged over blocks of 40~nm for the whole lens (5,000 blocks across), 20~nm for one quadrant of the lens and for the polarization-switched hologram, and 30~nm for the color hologram. The fill fraction of the binary layout is 0.442 for the lens within its circular aperture, 0.398 for the color hologram and 0.400 for the polarization-switched hologram.

\end{document}